\documentclass[runningheads,a4paper]{llncs}

\usepackage{amssymb}
\usepackage{xcolor}
\usepackage{epsfig}
\usepackage{multirow}
\usepackage{amsmath}
\usepackage{amsfonts}
\usepackage{graphics}
\usepackage{graphicx}
\usepackage{subfigure}
\usepackage{threeparttable}
\usepackage{subfig}
\usepackage{latexsym}
\usepackage{algorithm}
\usepackage{algpseudocode}
\usepackage{bbm}
\usepackage{algorithmicx}
\usepackage{geometry}
\usepackage{xcolor}

\newcommand*\swallow[1]{}
\algnewcommand\algorithmicinput{\textbf{Input:}}
\algnewcommand\INPUT{\item[\algorithmicinput]}
\algnewcommand\algorithmicoutput{\textbf{Output:}}
\algnewcommand\OUTPUT{\item[\algorithmicoutput]}

\newcommand{\quotes}[1]{``#1''}
\newcommand{\etal}{\textit{et al}.}

\definecolor{shadecolor}{RGB}{255,255,0}

\begin{document}

\mainmatter  

\title{COCACOLA: binning metagenomic contigs using sequence COmposition, read CoverAge, CO-alignment, and paired-end read LinkAge}


%
%
\author{Yang Young Lu$^{1}$%
\and Ting Chen$^{1,3}$\and Jed A. Fuhrman$^{2}$\and Fengzhu Sun$^{1,4*}$}
%


\institute{$^{1}$Molecular and Computational Biology Program, Department of Biological Sciences, University of
Southern California, CA, USA\\
$^{2}$Department of Biological Sciences and Wrigley Institute for Environmental Studies, University of Southern California, CA, USA\\
$^{3}$Center for Synthetic and Systems Biology, TNLIST, Beijing, China\\
$^{4}$Center for Computational Systems Biology, Fudan University, Shanghai, China}

%
%

\maketitle

\begin{abstract}
The advent of next-generation sequencing (NGS) technologies enables researchers to sequence complex microbial communities directly from environment. Since assembly typically produces only genome fragments, also known as contigs, instead of entire genome, it is crucial to group them into operational taxonomic units (OTUs) for further taxonomic profiling and down-streaming functional analysis. OTU clustering is also referred to as binning. We present COCACOLA, a general framework automatically bin contigs into OTUs based upon sequence composition and coverage across multiple samples.
The effectiveness of COCACOLA is demonstrated in both simulated and real datasets in comparison to state-of-art binning approaches such as CONCOCT, GroopM, MaxBin and MetaBAT. The superior performance of COCACOLA relies on two aspects. One is employing $L_{1}$ distance instead of Euclidean distance for better taxonomic identification during initialization. More importantly, COCACOLA takes advantage of both hard clustering and soft clustering by sparsity regularization. In addition, the COCACOLA framework seamlessly embraces customized knowledge to facilitate binning accuracy. In our study, we have investigated two types of additional knowledge, the co-alignment to reference genomes and linkage of contigs provided by paired-end reads, as well as the ensemble of both. We find that both co-alignment and linkage information further improve binning in the majority of cases. COCACOLA is scalable and faster than CONCOCT ,GroopM, MaxBin and MetaBAT.

The software is available at https://github.com/younglululu/COCACOLA
\end{abstract}

\section{Introduction}
Metagenomic studies aim to understand microbial communities directly
from environmental samples without cultivating member species
\cite{riesenfeld2004metagenomics}. The next-generation sequencing
technologies (NGS) allow biologists to extract genomic data with unprecedented high resolution and sufficient sequence depth, offering insights into complex microbial communities even including
species with very low abundance \cite{albertsen2013genome}. To
further investigate the taxonomic structure of microbial samples,
assembled sequence fragments, also known as contigs, need be grouped
into operational taxonomic units (OTUs) that ultimately represent genomes or significant parts of genomes. OTU clustering is also
called binning  (or genomic binning), serving as the key step towards taxonomic profiling
and downstream functional analysis. Therefore, accurate binning of
the contigs is an essential problem in metagenomic studies.

Despite extensive studies, accurate binning of contigs remains
challenging for several major reasons, including chimeric
assemblies due to repetitive sequence regions within or across
genomes, sequencing errors or artifacts, strain-level variation
within the same species, etc.
\cite{alneberg2014binning,mande2012classification} The currently
available binning methods can be broadly categorized into
classification and clustering approaches. Classification
approaches are \quotes{taxonomy dependent}, that is, reference
databases are needed for the assignment from contigs or reads to
meaningful taxons. The classification is either based on homology
due to sequence identity, or genomic signatures such as
oligonucleotide composition patterns and taxonomic clades.
Homology-based methods include MEGAN \cite{huson2007megan} that
assigns reads to the lowest common taxonomic ancestor. Examples of
genomic signature-based methods include PhyloPythia
\cite{mchardy2007accurate} and Kraken \cite{wood2014kraken} that are
composition-based classifiers and naive Bayesian classifier (NBC)
\cite{rosen2011nbc}, a clade-specific approach. In addition, hybrid
methods are available to take both alignment and composition-based
strategy into consideration, such as PhymmBL \cite{brady2009phymm}
and SPHINX \cite{mohammed2011sphinx}.

In comparison, clustering approaches are \quotes{taxonomy
independent}, that is, no additional reference databases or
taxonomic information is needed. These approaches require
similarity measurements from GC content, tetra-mer
composition
\cite{albertsen2013genome,chatterji2008compostbin,yang2010unsupervised}, 
or Interpolated Markov Models
\cite{kelley2010clustering}, 
to contig coverage profile \cite{baran2012joint,wu2011novel}.

{Recently, several methods have been
developed to bin contigs using the coverage profiles of the contigs
across multiple metagenomic samples \cite{albertsen2013genome,alneberg2014binning,carr2013reconstructing,imelfort2014groopm,kang2015metabat,nielsen2014identification,wu2015maxbin}
. Here the coverage of a contig
is defined as the fraction of reads mapped to the contig in a sample.
The idea is that if two contigs are from the same genome, their
coverage profiles across multiple samples should be highly
correlated. These methods can be further improved by integrating
coverage profiles with the sequence tetra-mer composition of
the contigs \cite{alneberg2014binning,imelfort2014groopm,kang2015metabat}}
\swallow{Recent trends tend to utilize coverage profiles across
multiple related samples . Based upon the assumption that contigs
with similar coverage profile tend to originate from the same
genome, superiority is denomstrated in combination of
tetra-mer composition and coverage profiles across multiple
related samples \cite{alneberg2014binning,imelfort2014groopm,kang2015metabat}.}
Among these methods, GroopM \cite{imelfort2014groopm} is
advantageous in its visualized and interactive pipeline. On one
hand, it is flexible, allowing users to merge and split bins under expert
intervention. On the other hand, in the absence of expert
intervention, the automatic binning results of GroopM is not as
satisfactory as CONCOCT \cite{alneberg2014binning}. CONCOCT
\cite{alneberg2014binning} makes use of the Gaussian mixture model
(GMM) to cluster contigs into bins. Also, CONCOCT provides a
mechanism to automatically determine the optimal OTU number by variational Bayesian model selection
\cite{corduneanu2001variational}. MetaBAT \cite{kang2015metabat} calculates integrated distance for pairwise contigs and then clusters contigs iteratively by modified K-medoids algorithm. And MaxBin \cite{wu2015maxbin} compares the distributions of distances between and within the same genomes.

In this paper we present COCACOLA, a general framework for contig
binning incorporating sequence \underline{CO}mposition, \underline{C}over\underline{A}ge, \underline{CO}-alignment and paired-end reads \underline{L}ink\underline{A}ge across multiple samples. By default, COCACOLA utilizes sequence
composition and coverage across multiple samples for binning. Compared to recent approaches such as CONCOCT, GroopM, MaxBin and MetaBAT, COCACOLA performs better in three aspects. Firstly, COCACOLA reveals superiority with respect to precision, recall and Adjusted Rand Index (ARI). Secondly, COCACOLA shows better robustness in the case of varying
number of samples. COCACOLA is scalable and faster than CONCOCT, GroopM, MaxBin andMetaBAT.

In addition, the COCACOLA framework seamlessly embraces customized knowledge to facilitate binning accuracy. In our study, we have investigated two types of knowledge, in particular, the co-alignment to reference genomes and linkage between contigs provided by paired-end reads. We find that both co-alignment and linkage information facilitate better binning performance in the majority of cases.

\section{Materials and Methods}
\label{methods}

\subsection{Problem Formulation}
\label{formulation}

A microbial community is comprised of a set of OTUs at different
abundance levels, and our objective is to put contigs into the genomic OTU bins from which they were originally derived. OTUs are expected to be disentangled based on contigs comprising either the discriminative abundance
or dissimilarity among sequences in terms of $l$-mer composition.
The rationale of binning contigs into OTUs relies on the underlying
assumption that contigs originating from the same OTU share similar
relative abundance as well as sequence composition.

Formally, we encode the abundance and composition of the $k$-th OTU  by a
$(M+V)$ dimensional feature vector, $W_{\cdot k}$, $k = 1,
2, \cdots, K$, where $M$ is the number of samples, $V$ is the number
of distinct $l$-mers, and $K$ is the total OTU number.
Specifically, $W_{m k}$ represents the abundance of the $k$-th OTU in the $m$-th sample
, $m = 1,2,\cdots ,M$, respectively. And $W_{M+v,k}$ stands for the
$l$-mer relative frequency composition of the $k$-th OTU, $v = 1,2,\cdots ,V$. Similarly, the feature vector of the $n$-th contig
is denoted as $X_{\cdot n}$. Let $\mathbbm{H}_{kn}$ be the
indicator function describing whether the $n$-th contig belongs to the $k$-th OTU,
i.e., $\mathbbm{H}_{kn}=1$ means the $n$-th contig originating from the $k$-th OTU
and $\mathbbm{H}_{kn}=0$ otherwise. Therefore, $X_{\cdot n}$ can be represented as:
\begin{equation}
X_{\cdot n} = \mathbbm{H}_{1n}W_{\cdot 1}+
\mathbbm{H}_{2n}W_{\cdot
2}+\cdots
+\mathbbm{H}_{kn}W_{\cdot K} ,\;\;n=1,2,\cdots,N \label{eq:form1}
\end{equation}
where $N$ is the number of contigs. Equation~(\ref{eq:form1}) can be further written into the matrix form:
\begin{equation}
X \approx W\mathbbm{H}  \; \; \; \; \; s.t. \; \;
W\geq 0,\;\; \mathbbm{H}\in \left \{ 0,1 \right \}^{K\times N}, \left
\|\mathbbm{H}_{\cdot n}\right \|_{0}=1
\label{eq:basicNMFL0}
\end{equation}
where $W=\left (W_{\cdot 1},W_{\cdot
2},\cdots ,W_{\cdot K} \right )$ is a $(M+V)\times K$
nonnegative matrix with each column encoding the feature vector of the corresponding OTU. And $\mathbbm{H}=\left ( \mathbbm{H}_{\cdot
1},\mathbbm{H}_{\cdot 2},\cdots ,\mathbbm{H}_{\cdot N} \right )$ is
a ${K\times N}$ binary matrix with each column encoding the
indicator function of the corresponding contig. $\left
\|\mathbbm{H}_{\cdot n}\right \|_{0} = \sum_{k = 1}^K \mathbbm{H}_{k
n} =1$ ensures the $n$-th contig belongs exclusively to only one
particular OTU.

The matrices $W$ and $\mathbbm{H}$ are obtained by
minimizing a certain objective function. In this paper we use Frobenius
norm, commonly known as the sum of squared error:
\begin{equation}
\arg \min _{W,\mathbbm{H}\geq 0}\left \|
X-W\mathbbm{H} \right \|_{F}^{2} \; \; \;
s.t. \; \;  \mathbbm{H}\in \left \{ 0,1 \right \}^{K\times N}, \left \|\mathbbm{H}_{\cdot n}\right
\|_{0}=1
\label{eq:basicNMFobjL0}
\end{equation}
Note that Equation~(\ref{eq:basicNMFobjL0}) is NP-hard by formulation as
an integer programming problem with an exponential number of
feasible solutions \cite{jiang2014clustering}. A common procedure to
tackle Equation~(\ref{eq:basicNMFobjL0}) relaxes binary constraint of
$\mathbbm{H}$ with numerical values. Hence Equation~(\ref{eq:basicNMFobjL0}) is reformulated as the following
minimization problem:
\begin{equation}
\arg \min _{W,H}\left \| X-WH \right \|_{F}^{2} \; \; \; \; \; s.t. \; \; W,H\geq 0
\label{eq:basicNMFobj}
\end{equation}
where $H$ serves as a coefficient matrix instead of an
indicator matrix. In the scenario of Equation~(\ref{eq:basicNMFobj}),
$W_{\cdot k}$, the feature vector of the $k$-th OTU , represents
the centroid of the $k$-th cluster. Meanwhile, each contig
$X_{\cdot n}$ is approximated by a weighted mixture of
clusters, where the weights are encoded in $H_{\cdot
n}$. In other words, relaxation of binary constraint makes the
interpretation from hard clustering to soft clustering,
{where hard clustering means that a
contig can be assigned to one OTU only, while soft clustering allows
a contig to be assigned to multiple OTUs}. It has been observed that
by imposing sparsity on each column of $H$, the hard
clustering behavior can be facilitated \cite{kim2008sparse}.
Therefore, Equation~(\ref{eq:basicNMFobj}) is further modified through
the Sparse Nonnegative Matrix Factorization (SNMF) form \cite{kim2008sparse}:
\begin{equation}
\arg \min _{W,H\geq 0}\left \| X-WH \right \|_{F}^{2}+\alpha \sum_{n=1}^{N}\left \|H_{\cdot n} \right \|_{1}^{2}
\label{eq:sparseNMFobj}
\end{equation}
where $\left \| \cdot   \right \|_{1}$ indicates $L_{1}$-norm. Due to
non-negativity of $H$, $\left \| H_{\cdot n}
\right \|_{1}$ stands for the column sum of the $n$-th column
vector of $H$. The parameter $\alpha > 0$ controls the
trade-off between approximation accuracy and the sparseness of
$H$. Namely, larger $\alpha$ implies stronger sparsity
while smaller value ensures better approximation accuracy.

\subsection{Feature Matrix Representation of Contigs}
\label{featureMat}
Similar to CONCOCT \cite{alneberg2014binning}, each contig longer
than 1000bp is represented by a $(M+V)$ dimensional column feature
vector including $M$ dimensional coverage and $V$ dimensional
tetra-mer composition. The coverage denotes the average
number of mapped reads per base pair from each of $M$ different
samples. While the tetra-mer composition denotes the tetra-mer frequency for the contig itself plus its reverse complement. Due
to palindromic tetra-mers, $V=136$.

Adopting the notation of CONCOCT \cite{alneberg2014binning}, the
coverage of all the $N$ contigs is represented by an $N \times M$
matrix $Y$, where $N$ is the number of contigs of interest and
$Y_{nm}$ indicates the coverage of the $n$-th contig from the $m$-th sample.
Whereas the tetra-mer composition of the $N$ contigs are
represented by an $N \times V$ matrix $Z$ where $Z_{nv}$ indicates
the count of $v$-th tetra-mer found in the $n$-th contig. Before
normalization, a pseudo-count is added to each entry of the coverage
matrix $Y$ and composition matrix $Z$, respectively. As for the
coverage, a small value is added, i.e., ${Y}'_{nm}=Y_{nm}+100/L_{n}$, analogous to a single read
aligned to each contig as prior, where $L_n$ is the length of the
$n$-th contig. As for the composition, a single count is simply
added, i.e., ${Z}'_{nv}=Z_{nv}+1$.

The coverage matrix $Y$ is firstly column-wise normalized (i.e., normalization within each individual sample), followed by row-wise
normalization (i.e., normalization across $M$ samples) to obtain
coverage profile $p$. The row-wise normalization aims to
mitigate sequencing efficiency heterogeneity among contigs.
\begin{equation}
{Y}''_{nm}=\frac{{Y}'_{nm}}{\sum_{n=1}^{N}{Y}'_{nm}} \\ \; \; \; \;
\; \; \; \;
p_{nm}=\frac{{Y}''_{nm}}{\sum_{m=1}^{M}{Y}''_{nm}}
\label{eq:coverageProfile}
\end{equation}
The composition matrix $Z$ is row-wise normalized for each contig (i.e., normalization across $M$ tetra-mer count) to obtain composition profile $q$:
\begin{equation}
q_{nv}=\frac{{Z}'_{nv}}{\sum_{v=1}^{V}{Z}'_{nv}}
\label{eq:compositionProfile}
\end{equation}
The feature matrix of contigs is denoted as $X=\left [
p\; q \right ]^{T}$, as the combination of
coverage profile $p$ and composition profile $q$.
To be specific, $X$ is a $(M+V)\times N$ nonnegative matrix
of which each column represents the feature vector of a particular
contig.

\subsection{Incorporating Additional Information into Binning}
\label{additionInfo}
We consider two types of additional knowledge that may enhance
the binning accuracy \cite{basu2008constrained}. One option is
paired-end reads linkage. Specifically, a high number of links connecting two
contigs imply high possibility that they belong to the same OTU.
Since the linkage may be erroneous due to the existence of
chimeric sequences, we keep linkages that are reported through
multiple samples. The other option is co-alignment to reference genomes. That is, two contigs mapped to the same reference genome support the evidence that they belong to the same OTU.

We encode additional knowledge by an undirected network in the form
of a nonnegative weight matrix $A$, where $A_{n
n'}$ quantifies the confidence level we believe the $n$-th contig and the $n'$-th
contig to be clustered together. Based upon the aforementioned matrix $\mathbf{A}$, a network regularization item is
introduced to measure the coherence of binning \cite{cai2011graph}:
\begin{equation}
\label{eq:graphReg}
R_{g}= \frac{1}{2}\sum_{n,n'=1}^{N}\left \| H_{\cdot n}-H_{\cdot n'} \right \|^{2}A_{nn'}
 = Tr(HLH^{T})
\end{equation}
where $Tr(\cdot )$ indicates the matrix trace, {the sum of items along the
diagonal}. $D$ denotes the diagonal matrix whose entries are column
sums (or row sums due to symmetry) of $A$, i.e.,
$D_{n n}=\sum_{n'=1}^{N}A_{n n'}$. The
$\mathit{Laplacian}$ $\mathit{matrix}$ \cite{chung1997spectral} is
defined as $L=D-A$. With convention we
use $\mathit{normalized}$ $\mathit{Laplacian}$ $\mathit{matrix}$
instead, that is,
$\mathcal{L}=D^{-1/2}LD^{-1/2}=I-D^{-1/2}AD^{-1/2}
\triangleq I-\mathcal{A}$.
By incorporating the network regularization in Equation~(\ref{eq:graphReg}), the objective function in Equation~(\ref{eq:sparseNMFobj}) changes to the following form:
\begin{equation}
\arg \min _{W,H\geq 0}\left \| X-WH \right \|_{F}^{2}+
\alpha \sum_{n=1}^{N}\left \| H_{\cdot n} \right \|_{1}^{2} +  \beta \, Tr(H\mathcal{L}H^{T})
\label{eq:finalObj}
\end{equation}
where the parameter $\beta > 0$ controls the trade-off of belief
between unsupervised binning and additional knowledge.
Namely, large $\beta$ indicates strong confidence on the
additional knowledge. Conversely, small $\beta$ puts more
weight on the data.

To utilize multiple additional knowledge sources together, a combined $\mathit{Laplacian}$ $\mathit{matrix}$ is constructed as a weighted average of individual $\mathit{Laplacian}$ $\mathit{matrices}$ $\bar{\mathcal{L}}=\left ( \sum_{d}\alpha _{d}\mathcal{L}_{d} \right )/\left ( \sum_{d}\alpha _{d} \right )$ where each positive weight $\alpha _{d}$ reflects the contribution of the corresponding information. For simplicity weights are treated equally in the paper.

\subsection{Optimization by Alternating Nonnegative Least Squares (ANLS)}
\label{anlsOptimzation}
Among comprehensive algorithms to solve Equation~(\ref{eq:finalObj}), the
multiplicative updating approach \cite{lee1999learning} is most
widely used. Despite its simplicity in implementation, slow
convergence is of high concern. This paper adopts a more efficient algorithm with
provable convergence called alternating nonnegative least squares (ANLS)
\cite{kim2008sparse}. 
ANLS iteratively handles two nonnegative least square (NNLS)
subproblems in Equation~(\ref{eq:anls}) until convergence. The ANLS
algorithm is summarized in Algorithm \ref{alg:anls}.
\begin{subequations}
\label{eq:anls}
\begin{align}
H\leftarrow & \arg \min _{H\geq 0}\left \| X-WH \right \|_{F}^{2}+\alpha \sum_{n=1}^{N}\left \| H_{\cdot n} \right \|_{1}^{2} + \beta \, Tr(H\mathbf{\mathcal{L}}H^{T}) \label{eq:anlsSub1} \\
W\leftarrow & \arg \min _{W\geq 0}\left \| X^{T}-H^{T}W^{T} \right \|_{F}^{2} \label{eq:anlsSub2}
\end{align}
\end{subequations}
\begin{algorithm}
\caption{Optimization by Alternating Nonnegative Least Squares}
\label{alg:anls}
\begin{algorithmic}[1]
\INPUT feature matrix $X\in \mathbb{R}^{(M+V)\times N}$, initial basis matrix $W\in \mathbb{R}^{(M+V)\times K}$ and coefficient matrix $H\in \mathbb{R}^{K \times N}$, tolerance threshold $\varepsilon$, maximum iteration threshold $T$ \Repeat
    \State Obtain optimal $H$ of Equation~(\ref{eq:anlsSub1}) by fixing $W$
    \State Obtain optimal $W$ of Equation~(\ref{eq:anlsSub2}) by fixing $H$
\Until{A particular stopping criterion involving $\varepsilon$ is satisfied or iteration number exceeds $T$}
\Statex
\OUTPUT $W$,$H$
\end{algorithmic}
\end{algorithm}
We solve Equation~(\ref{eq:anlsSub1}) by block coordinate descent (BCD), that is, we divide Equation~(\ref{eq:anlsSub1}) into $N$ subproblems and minimize the objective function with respect to each subproblem at a time while keep the rest fixed:
\begin{equation}
\label{eq:anlsHSub}
\begin{aligned}
& \arg \min _{H_{\cdot n}\geq 0}  \left \| X_{\cdot n}-WH_{\cdot n} \right \|_{2}^{2}+\alpha \left \| H_{\cdot n} \right \|_{1}^{2}+\beta \,H_{\cdot n}^{T}\mathcal{L}H_{\cdot n}, \; n=1,\cdots,N \\
= & \arg \min _{H_{\cdot n}\geq 0}  \left \| X_{\cdot n}-WH_{\cdot n} \right \|_{2}^{2}+\alpha \left \| H_{\cdot n} \right \|_{1}^{2}+\beta \,H_{\cdot n}^{T}(H_{\cdot n}-2\sum_{n'=1}^{N}\mathcal{A}_{nn'}H_{\cdot n'}^{old}) \\
= & \arg \min _{H_{\cdot n}\geq 0}  \left \| X_{\cdot n}-WH_{\cdot n} \right \|_{2}^{2}+\alpha \left \| H_{\cdot n} \right \|_{1}^{2}+\beta \,\left \| H_{\cdot n}-\sum_{n'=1}^{N}\mathcal{A}_{nn'}H_{\cdot n'}^{old} \right \|_{2}^{2}
\end{aligned}
\end{equation}
where the matrix $H^{old}$ denotes the value of $H$ obtained from the previous iteration. Following Jacobi updating rule, we combine $N$ subproblems in Equation~(\ref{eq:anlsHSub}) into the matrix form:
\begin{equation}
\label{eq:anlsH}
\arg \min _{H\geq 0} \left \| \begin{pmatrix}
X\\
0_{1\times N}\\
\sqrt{\beta }H^{old}\mathcal{A}
\end{pmatrix}-\begin{pmatrix}
W\\
\sqrt{\alpha }e_{1\times K}\\
\sqrt{\beta }I_{K}
\end{pmatrix}H \right \|_{F}^{2}
\end{equation}
where $0_{1\times N}$ is a $N$ dimensional row vector of all 0, $e_{1\times K}$ is a $K$ dimensional row vector of all 1.
\subsection{Initialization of $W$ and $H$}
\label{initialization}
Note that we need to initialize $W$ and $H$ as the input to
Algorithm \ref{alg:anls}. A good initialization not only enhances
the accuracy of the solution, but facilitates fast convergence to a better local minima as well \cite{langville2006initializations}.
We initialize $W$ and $H$ by K-means clustering, namely, $W$ is set to be the K-means centroid of $X$ with each column $W_{\cdot k}$ corresponding to the feature vector of the $k$-th centroid. Meanwhile, $H$ is set to be the indicator matrix encoding the cluster assignment.

The distance measurement contributes crucially to the success of binning. Ideally, a proper distance measurement exhibits more distinguishable taxonomic difference. The traditional K-means approach takes Euclidean distance as default measurement to quantify closeness. However, 
as for the coverage profile, Su \etal \cite{su2012impact} shows $L_{1}$ distance produces more reasonable binning results than Euclidean and correlation-based distances. As for the composition profile, $L_{1}$ distance also reveals superiority over Euclidean and cosine distances \cite{liao2014new}. Therefore, our method adopts K-means clustering with $L_{1}$ distance. Once preliminary K-means clustering is achieved, we eliminate suspicious clusters with few contigs using the bottom-up L Method \cite{salvador2004determining}. 
\subsection{Parameter Tuning}
\label{paramTuning}
We have two parameters $(\alpha ,\beta )$ to be tuned in our
algorithm. Traditional cross-validation-like strategy demands
searching through a two dimensional grid of candidate values, which is
computationally unaffordable in the case of large datasets. Instead,
we firstly search a good marginal $\alpha$ value by fixing
$\beta=0$. After that, a one dimensional search is performed on a
range of candidate $\beta$ values while keeping $\alpha$ fixed.

In our implementation, when $\beta=0$, $\alpha$ is approximated by the Lagrange Multiplier of Equation~(\ref{eq:basicNMFobj}) with constraint $\sum_{n=1}^{N}\left ( \left \| H_{\cdot n} \right \|_{1}-1 \right )^{2}=0$, denoted by $\alpha^{\ast }$. Then we run the algorithm with respect to each candidate $\beta$ and fixed $\alpha=\alpha^{\ast }$, resulting in corresponding binning results with various cluster number. Notice that traditional internal cluster validity indices are only applicable on the basis of fixed cluster number scenario \cite{wiwie2015comparing}, such as SSE (Sum of Square Error), Davies-Bouldin index \cite{davies1979cluster}, 
etc. To be specific, the indices have the tendency towards monotonically increase or decrease as the cluster number increases \cite{liu2013understanding}. We tackle the impact of monotonicity by adopting TSS minimization index \cite{tang2005improved}, that is, we choose the candidate $\beta$ with minimum TSS value, recorded as $\beta^{\ast }$. Then we can solve Equation~(\ref{eq:finalObj}) by using $(\alpha^{\ast } ,\beta^{\ast} )$ as selected regularization parameters.
\subsection{Post-processing}
\label{postprocess}
The resulting binning obtained from Algorithm \ref{alg:anls} may contain clusters that are closely mixed to each other. Therefore, we define \textit{separable conductance} as an effective measurement to diagnose the coupling closeness of pairwise clusters, so as to determine whether to merge them. Namely, we consider each cluster as having a spherical scope centered at its centroid. To be robust against outliers, the radius is chosen as the third quartile among the intra-cluster distances. The \textit{separable conductance} between the $c_{1}$-th cluster and the $c_{2}$-th cluster, $sep(c_{1},c_{2})$, is defined as the number of contigs from the $c_{1}$-th cluster also included in the spherical scope of the $c_{2}$-th cluster, divided by the smaller cluster size of two. Intuitively, the \textit{separable conductance} exploits the overlap between two clusters. The procedure of post-processing works as follows: we keep picking the pair of clusters with maximum \textit{separable conductance} and merge them until it fails to exceed a certain threshold. The threshold is set to be 1 in this study.
\subsection{Datasets}
\label{datasets}

Alneberg \etal \, \cite{alneberg2014binning} simulated a \quotes{species} dataset and another \quotes{strain} dataset. Both simulated datasets were constructed based upon 16S rRNA samples originated from the Human Microbiome Project (HMP) \cite{human2012structure}. The relative abundance profiles of the different species/strains for the simulation were based on the HMP samples as well.

The simulated \quotes{species} dataset consisted of 101 different species across 96 samples. It aimed to test the ability of CONCOCT to cluster contigs in complex populations \cite{alneberg2014binning}. The species were approximated by the operational taxonomic units (OTUs) from HMP with more than 3\% sequence differences. Each species was guaranteed to appear in at least 20 samples. A total of 37,628 contigs remain for binning after co-assembly and filtering.

The simulated \quotes{strain} dataset aimed to test the ability of CONCOCT to cluster contigs at different levels of taxonomic resolution \cite{alneberg2014binning}. To be more specific, the simulated \quotes{strain} dataset consisted of 20 different species or strains from the same species across 64 samples, including five different \textit{E. coli} strains, five different \textit{Bacteroides} species, five different species from different \textit{Clostridium} genera, and five different \textit{gut} bacteria. It was challenging for CONCOCT to separate the five different \textit{E. coli} strains \cite{alneberg2014binning}. A total of 9,417 contigs remain for binning after co-assembly and filtering.

In addition to two simulated datasets, We use a time-series study of 11 fecal microbiome samples from a premature infant \cite{sharon2013time}, denoted as the \quotes{Sharon} dataset. Since the true species that contigs belong to are not known, we assign the class labels by annotating contigs using the TAXAassign script \cite{taxaassign}. As a result, $2,614$ out of $5,579$ contigs are unambiguously labeled on the species level for evaluation. Another real dataset embody $264$ samples from the MetaHIT consortium \cite{qin2010human} (SRA:ERP000108), the same dataset used in MetaBAT \cite{kang2015metabat}, denoted as the \quotes{MetaHIT} dataset. $17,136$ out of $192,673$ co-assembled contigs are unambiguously labeled on the species level for evaluation.
\subsection{Evaluation Criteria}
\label{evalCriteria}
To evaluate a binning result with $K$ clusters symbolizing predicted OTUs, against targeted $S$ species, a $K \times S$ matrix $A = (a_{ks})$ can be constructed so that $a_{ks}$ indicates the shared number of contigs between the $k$-th OTU and the $s$-th species. Therefore, $a_{k\cdot}=\sum_{s} a_{ks}$ and $a_{\cdot s}=\sum_{k} a_{ks}$ stand for the size of the $k$-th OTU and the $s$-th species, respectively.

One evaluation measurement focuses on if pairs of contigs belonging to the same species can be clustered together  
, such as Adjusted Rand Index (ARI)
. Given the knowledge of the species to which each contig is mapped, we treat the corresponding species as class labels. Then the classification of pairs of contigs falls into one of the four cases: $TP$ (True Positive) and $FP$ (False Positive) represent the number of pairs of contigs that truly belong to the same species being clustered into the same OTUs and distinct OTUs, respectively; $FN$ (False Negative) and $TN$ (True Negative) stand for the number of pairs of contigs from different species being clustered into the same OTUs and distinct OTUs, respectively. ARI is calculated as Equation~(\ref{eq:ari}):
\begin{equation}
\begin{aligned}
ARI & =\frac{2\left ( TP\times TN- FP\times FN\right )}{FP^{2}+FN^{2}+2\times TP\times TN +\left ( TP+TN \right )\times \left ( FP+FN \right )} \\
 & =\frac{\sum_{k,s}\binom{a_{ks}}{2}-t_{3}}{\frac{1}{2}\left ( t_{1}+t_{2} \right )-t_{3}} \\
 &where \; \;  t_{1}=\sum_{k}\binom{a_{k\cdot}}{2}, \; \;  t_{2}=\sum_{s}\binom{a_{\cdot s}}{2}, \; \; t_{3}=\frac{2t_{1}t_{2}}{\binom{N}{2}}
\label{eq:ari}
\nonumber
\end{aligned}
\end{equation}
where $N$ is the total number of contigs. In addition to ARI, we also consider \emph{precision} and \emph{recall} defined as follows:
\[ \texttt{precision} = \frac{1}{N}\sum_{k}\max_{s}\left \{ a_{ks} \right \}. \; \; \; \; \; \; \texttt{recall} = \frac{1}{N}\sum_{s}\max_{k}\left \{ a_{ks} \right \}. \]

\section{Results}
\label{results}
Given the same input, i.e., sequence composition and coverage across multiple samples, we show the effectiveness of COCACOLA on simulated \quotes{species} and \quotes{strain} datasets, in comparison with three state-of-art, methodologically distinct methods for contigs binning: CONCOCT \cite{alneberg2014binning}, GroopM \cite{imelfort2014groopm}, MaxBin \cite{wu2015maxbin} and MetaBAT \cite{kang2015metabat}. The comparison excludes Canopy \cite{nielsen2014identification} that is based on binning co-abundant gene groups instead of binning contigs. 
Furthermore, we investigate the performance improvement of COCACOLA after incorporating two additional knowledge, co-alignment to reference genomes and linkage between contigs provided by paired-end reads, as well as the ensemble of both. Results reveal both information facilitating better performance in the majority of cases.
Finally, we report the performance of COCACOLA on two real datasets.

\subsection{Performance on the Simulated Datasets}
Even though both COCACOLA and CONCOCT are able to deterine the OTU number automatically, an initial estimation of OTU number $K$ is needed to start from. Since the OTU number is usually unknown, we study the binning performance with respect to the value of $K$ chosen empirically, then we study the effect of varying $K$ on the binning results (Section \ref{resultCompK}).

We observed that K-means clustering tends to generate empty clusters given large $K$. Our strategy is to increase $K$ until there are more than $K/2$ empty clusters, and we choose the corresponding $K$ as the input. At this stage, we emphasize more on the redundancy of OTU number rather than the accuracy. Thus, we obtain $K=192$ and $K=48$ as input to the simulated \quotes{species} and \quotes{strain} dataset, respectively.

\begin{figure}
\centering
\begin{tabular}{cc}
\includegraphics[width=0.35\textwidth]{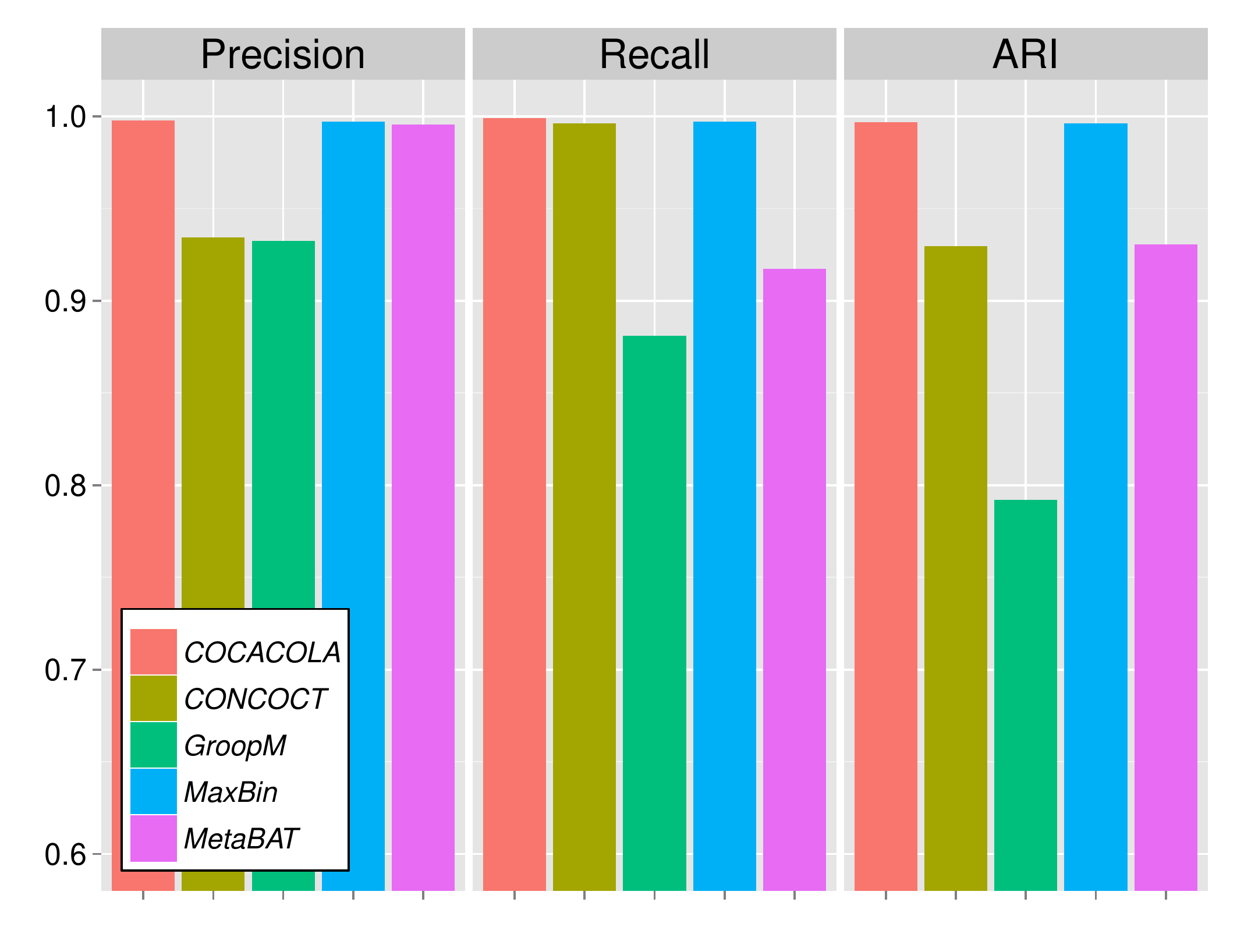} &   \includegraphics[width=0.35\textwidth]{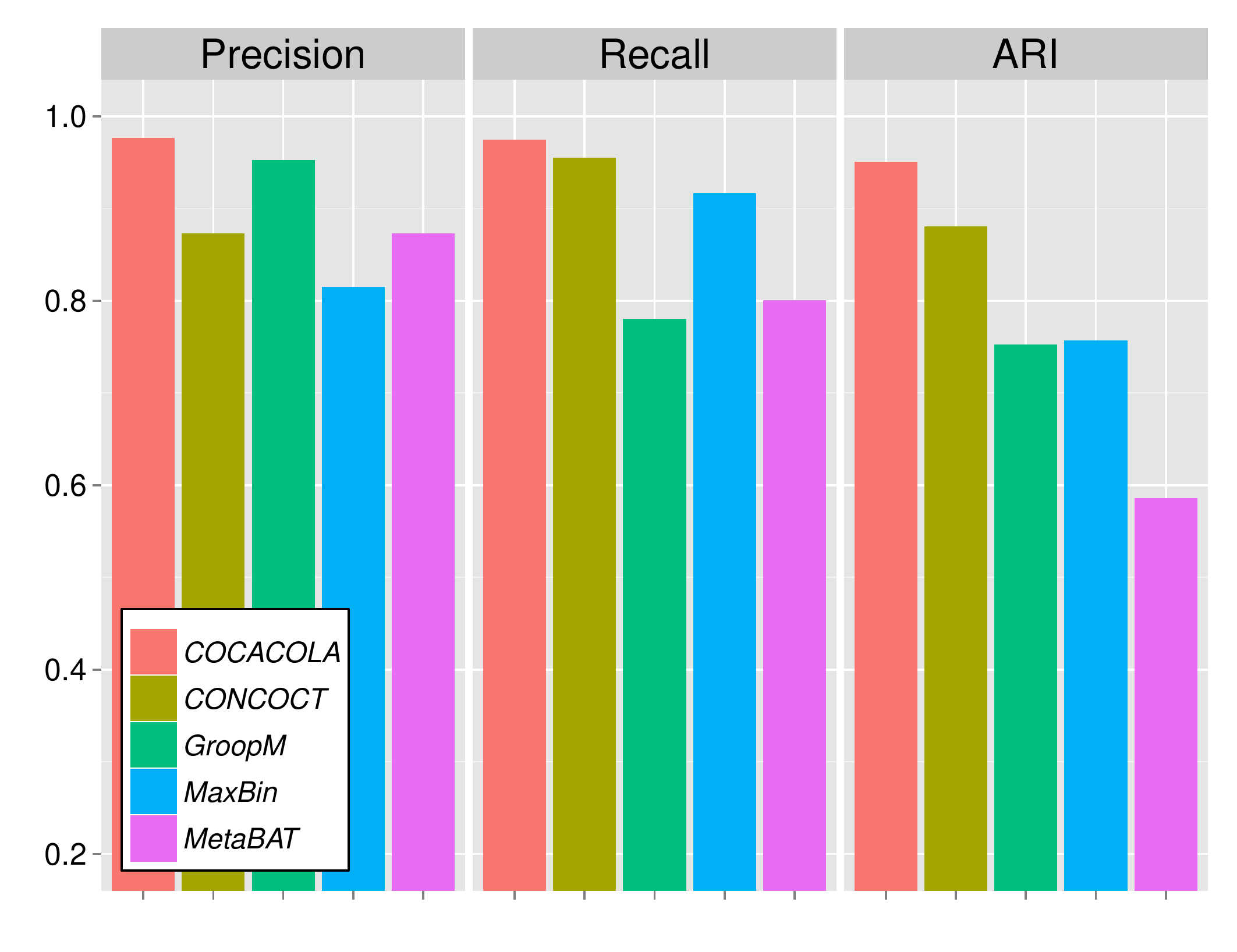} \\
(a) simulated \quotes{species} dataset & (b) simulated \quotes{strain} dataset \\
\includegraphics[width=0.35\textwidth]{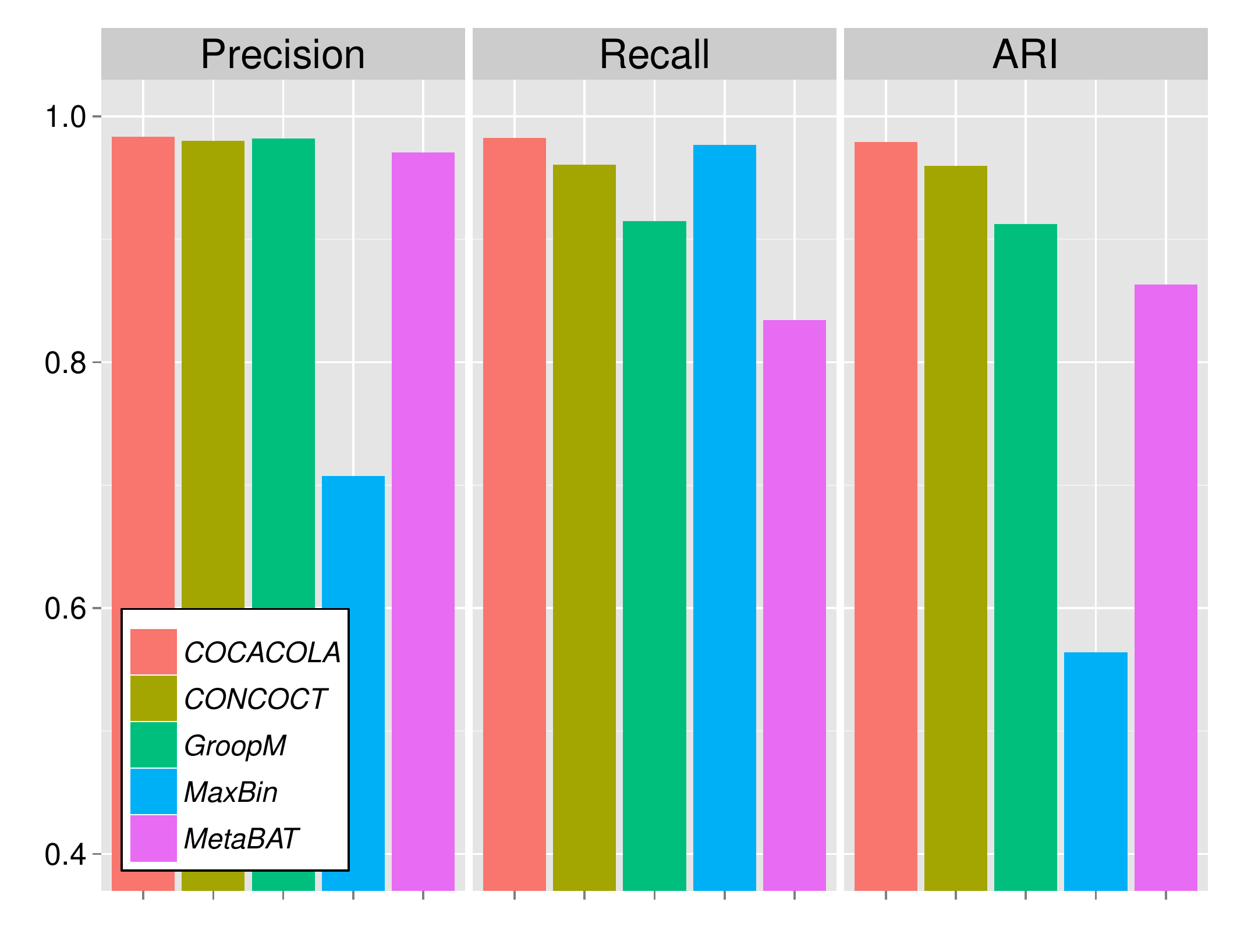} &
\includegraphics[width=0.35\textwidth]{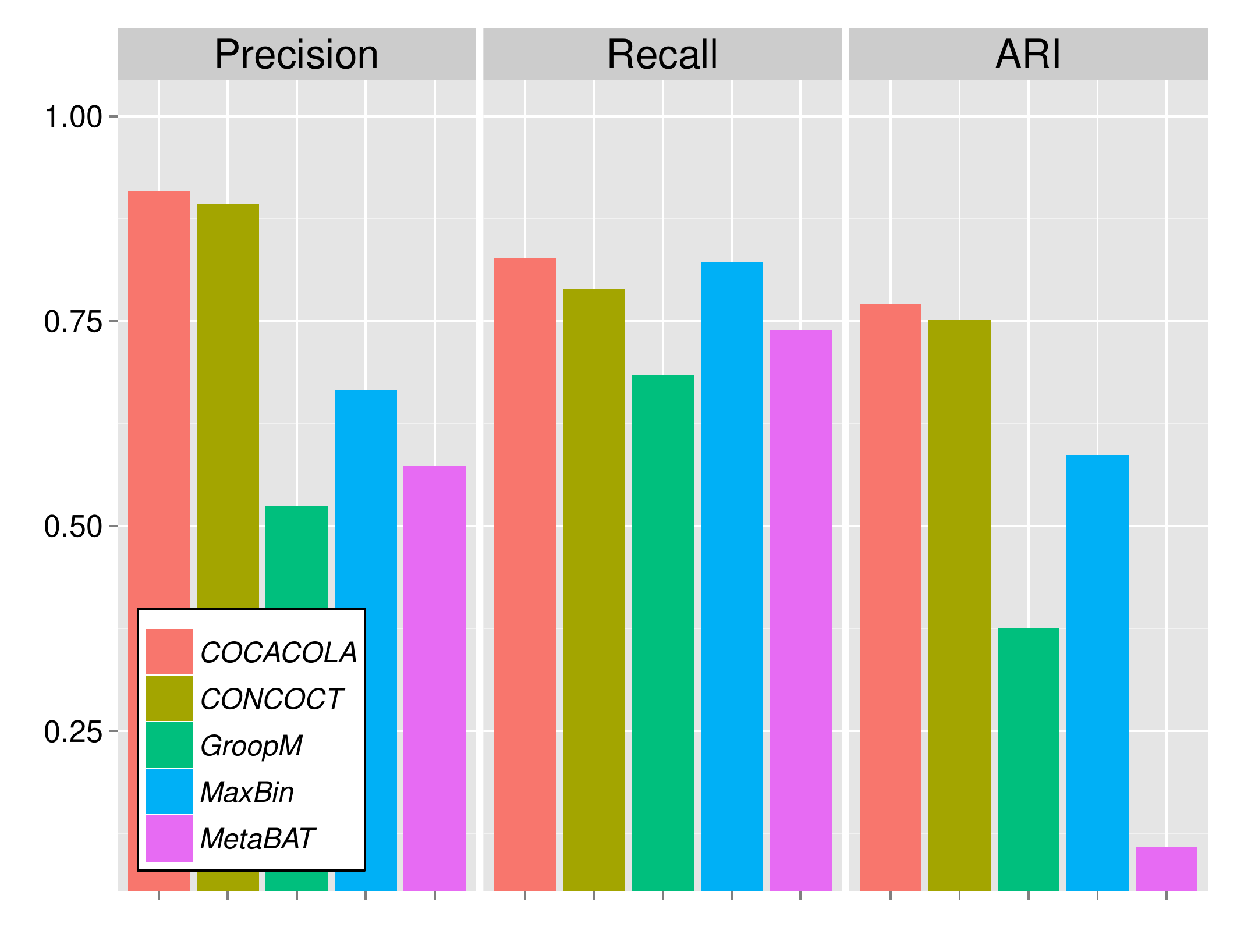} \\
(c) real \quotes{Sharon} dataset & (d) real \quotes{MetaHIT} dataset \\
\end{tabular}
\caption{The performance of COCACOLA, CONCOCT, GroopM, MaxBin and MetaBAT on both simulated datasets (a,b) and real datasets (c,d). }
\label{fig:autoK}
\end{figure}

For the simulated \quotes{species} dataset, Figure~\ref{fig:autoK}(a) compares COCACOLA against CONCOCT, GroopM, MaxBin and MetaBAT in terms of precision, recall and ARI. The precision of COCACOLA is $0.9978$, suggesting that almost all contigs within each cluster originate from the same species. In comparison,  the precision of CONCOCT, GroopM, MaxBin and MetaBAT is $0.9343$, $0.9324$, $0.9973$ and $0.9958$, respectively. The recall obtained by COCACOLA is $0.9993$, implying that nearly all contigs derived from the same species are grouped into the same clusters. In contrast, the recall of CONCOCT, GroopM, MaxBin and MetaBAT is $0.996$, $0.881$, $0.9973$ and $0.9174$, respectively. As for ARI, COCACOLA achieves $0.997$ while CONCOCT, GroopM, MaxBin and MetaBAT get $0.9296$, $0.7922$, $0.9961$ and $0.9308$, respectively.

For the simulated \quotes{strain} dataset, the results are shown by Figure~\ref{fig:autoK}(b). The precision, recall and ARI of COCACOLA reach $0.9766$, $0.9747$ and $0.9512$, respectively. In comparison, CONCOCT, GroopM, MaxBin and MetaBAT achieve $0.8733$, $0.9525$, $0.8151$ and $0.8730$ in terms of precision, $0.9552$, $0.7805$, $0.9167$ and $0.8009$ in terms of recall, $0.8809$, $0.7529$, $0.757$ and $0.5858$ in terms of ARI, respectively.

We conclude that COCACOLA performs well in constructing species from highly complicated environmental samples. Besides, COCACOLA performs well in handling strain-level variations, which cannot be fully resolved due to assembly limitation \cite{alneberg2014binning}.
\subsection{Impact of Varying Number of Clusters $K$}
\label{resultCompK}

\begin{figure}
\centering
\begin{tabular}{ccc}
\includegraphics[width=0.32\textwidth]{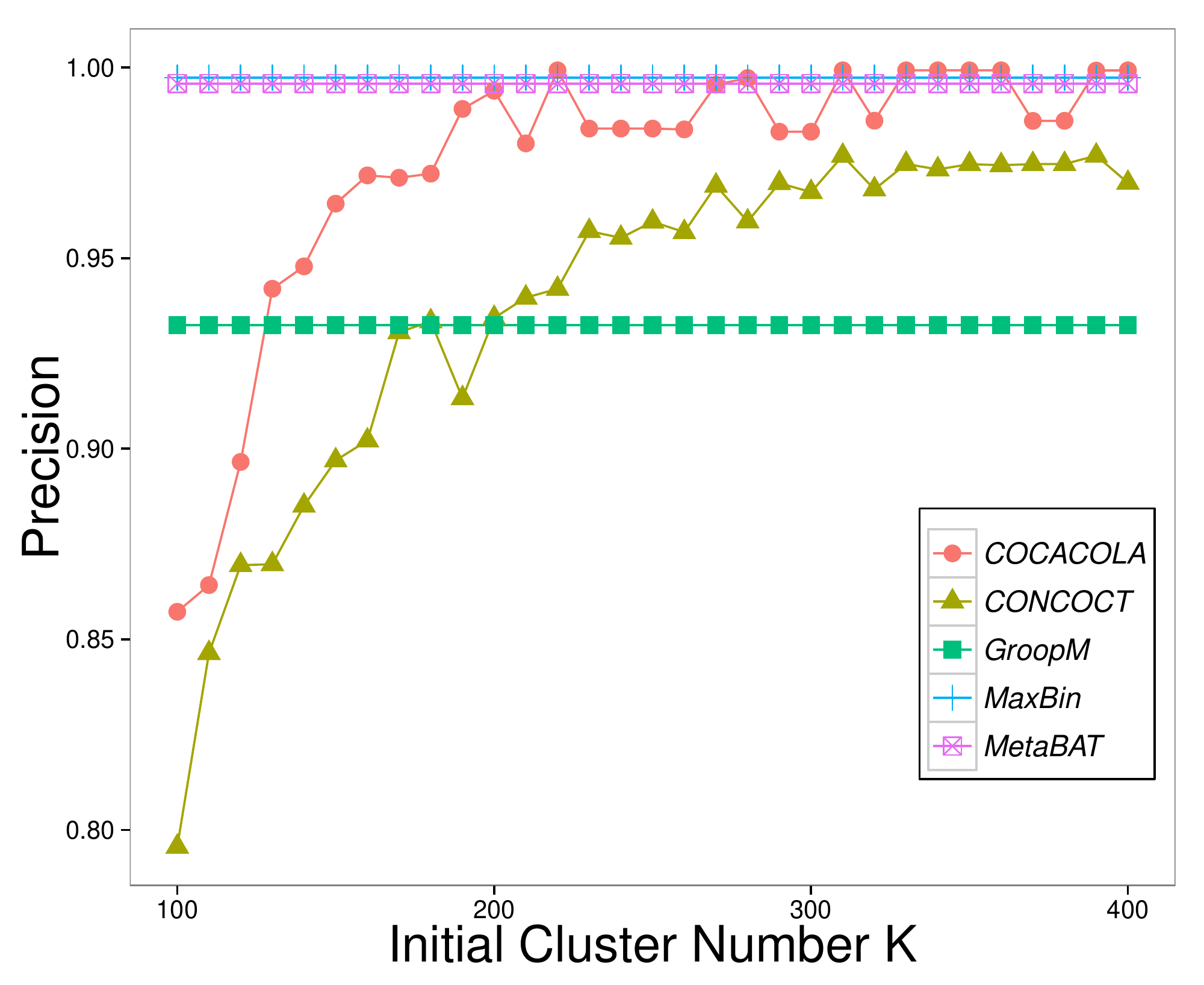} &   \includegraphics[width=0.32\textwidth]{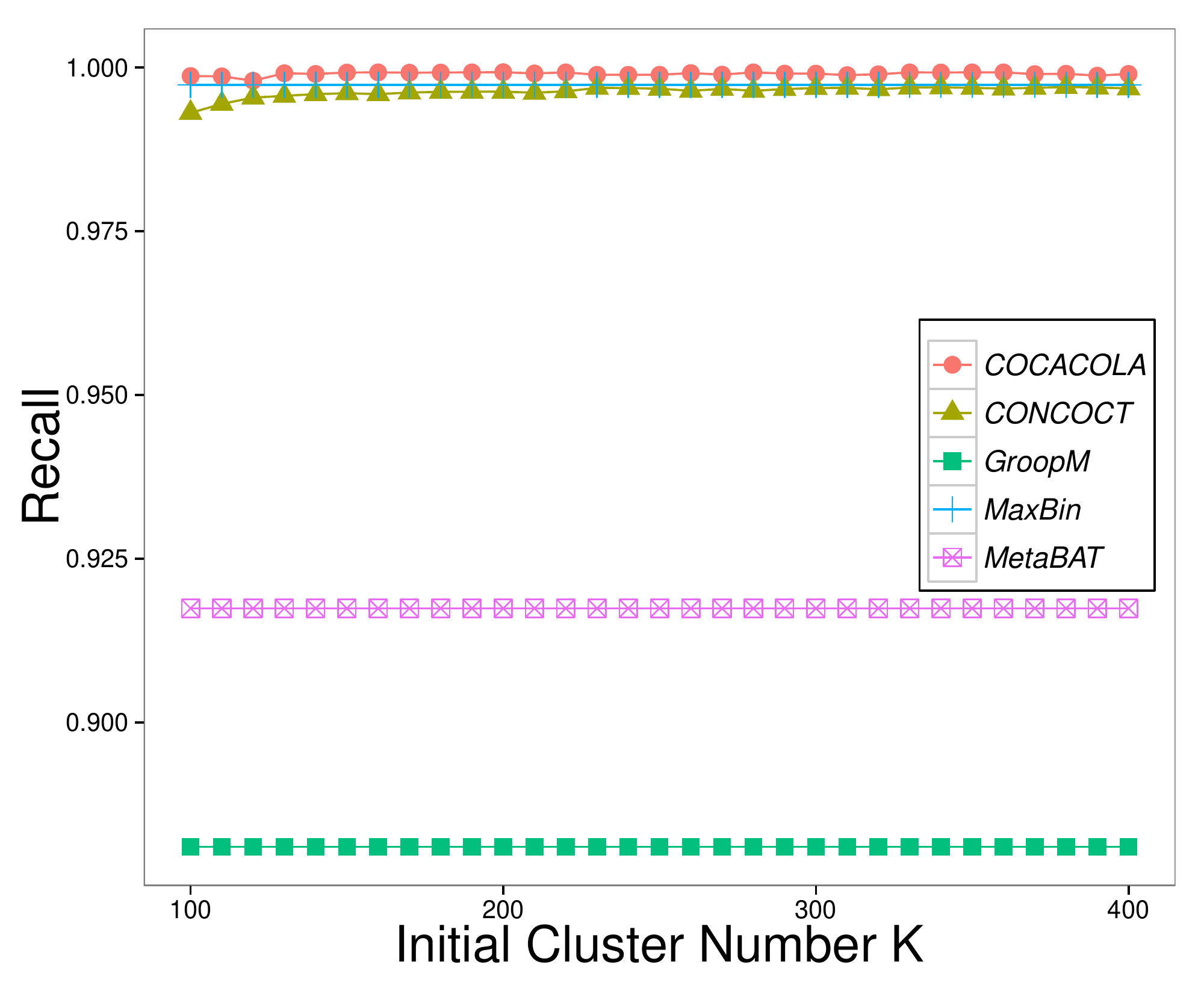} &   \includegraphics[width=0.32\textwidth]{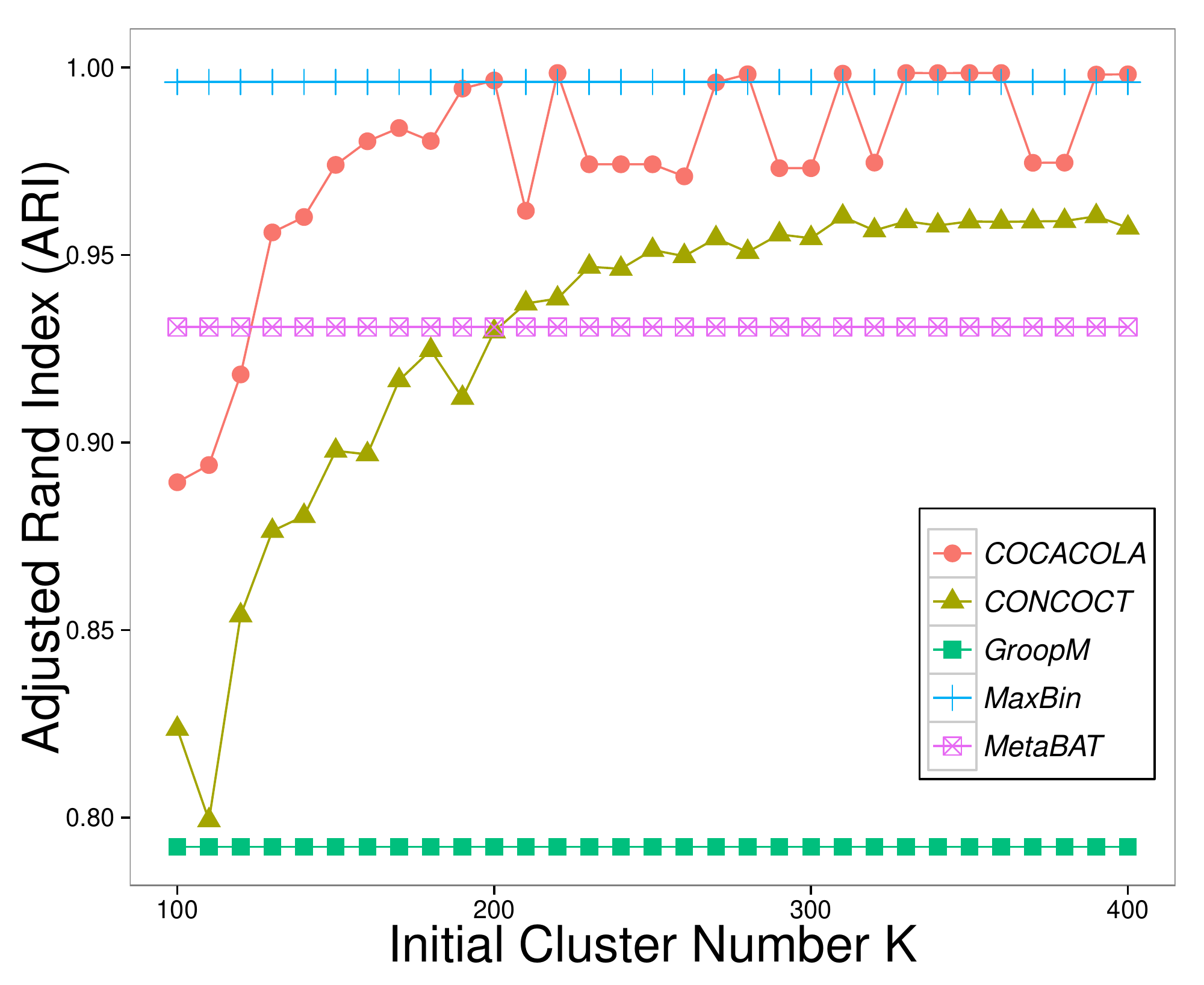} \\
(a) & (b) & (c)\\
\includegraphics[width=0.32\textwidth]{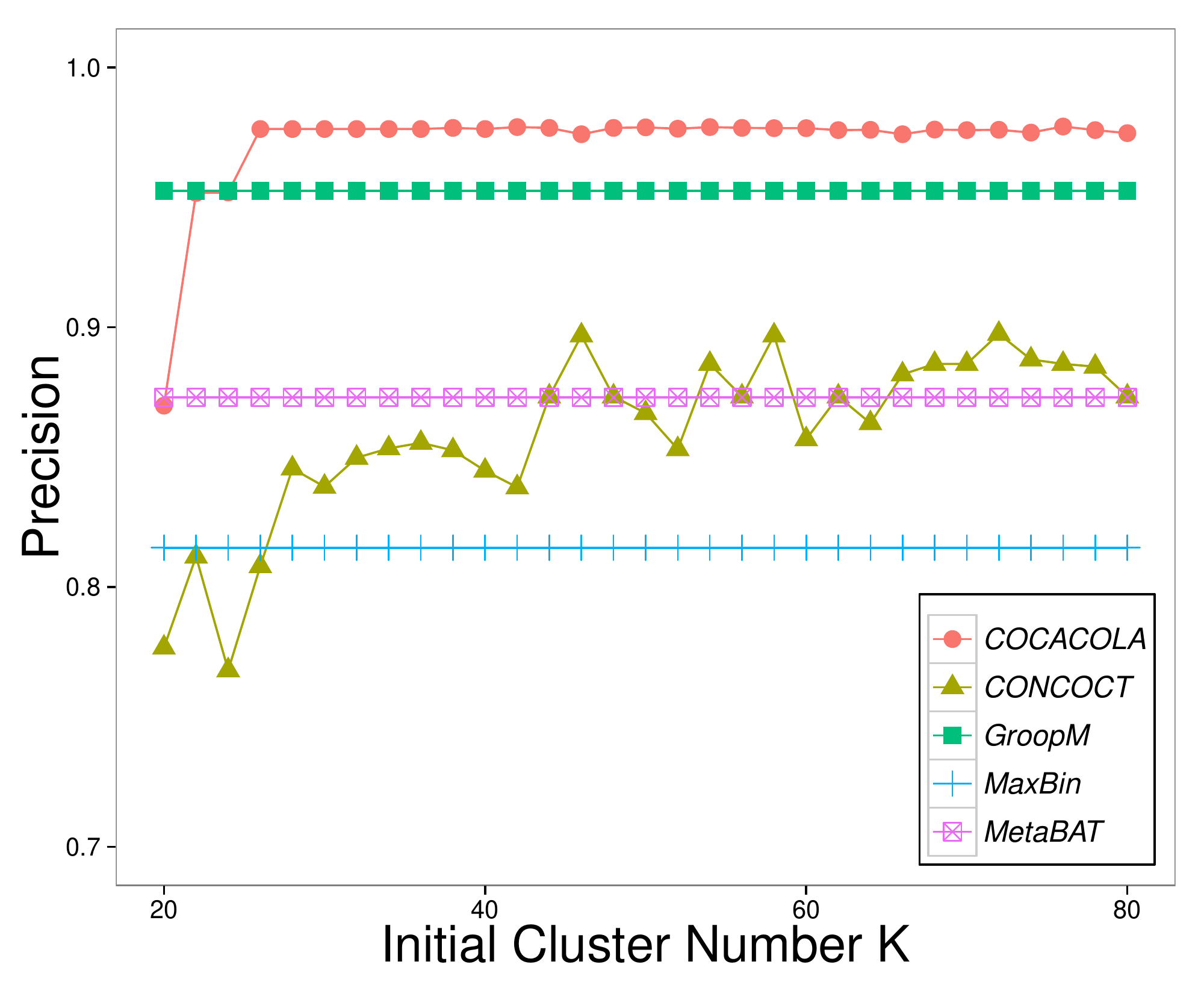} &   \includegraphics[width=0.32\textwidth]{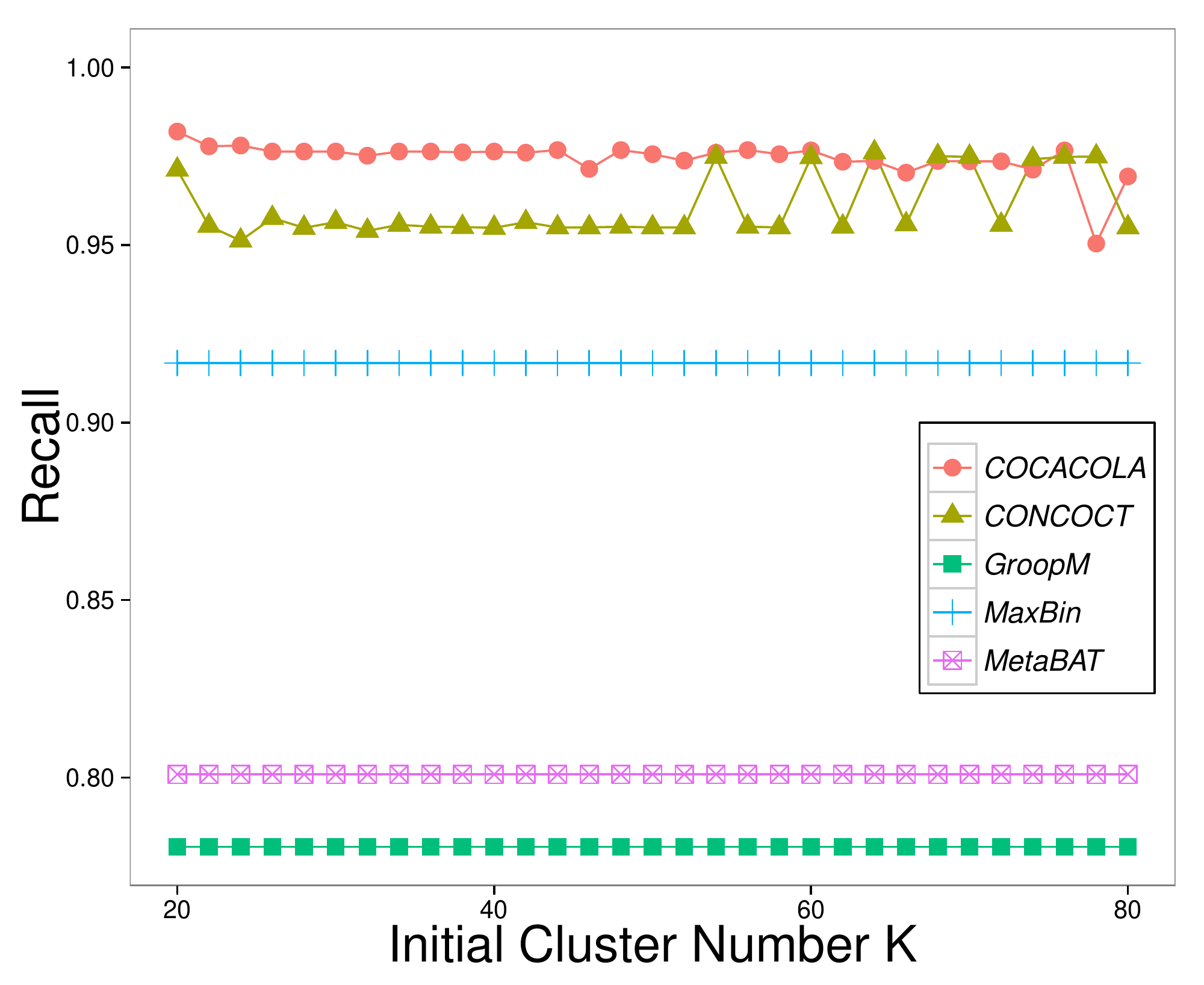} &   \includegraphics[width=0.32\textwidth]{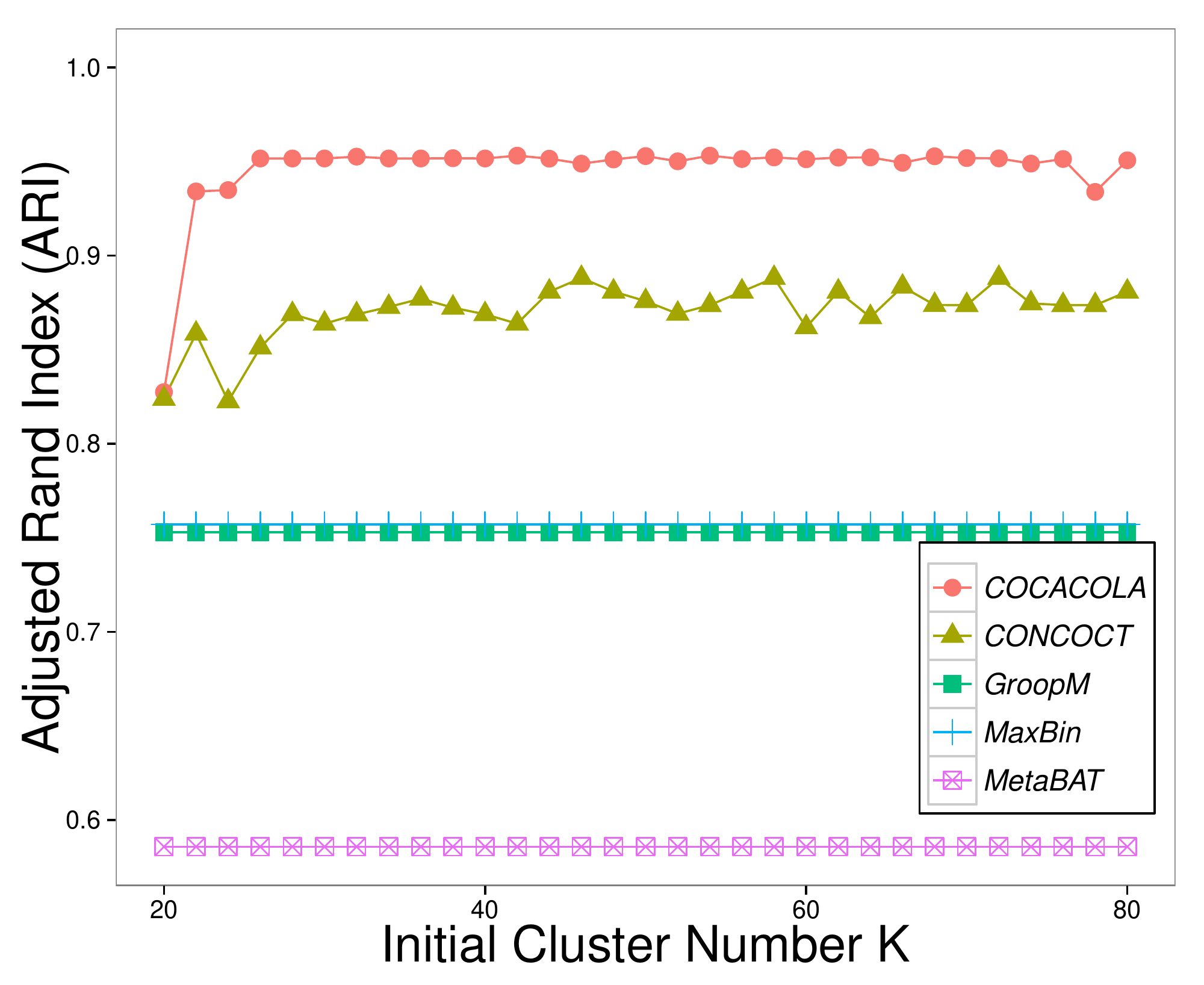} \\
(d) & (e) & (f)\\
\end{tabular}
\caption{The performance of COCACOLA, CONCOCT, GroopM, MaxBin and MetaBAT in terms of precision, recall and ARI on the simulated \quotes{species} and \quotes{strain} datasets. The results of simulated \quotes{species} dataset are depicted by (a)-(c). And the results of simulated \quotes{strain} dataset are depicted by (d)-(f). }
\label{fig:manualK}
\end{figure}

We study the impact of varying $K$ on the binning results. We apply COCACOLA and CONCOCT to both simulated \quotes{species} and \quotes{strain} datasets under different values of $K$ (Figure~\ref{fig:manualK}). We let $K$ range from the exact number of the real species to four times of it, with 10\% increment. Since GroopM, MaxBin  and MetaBAT doesn't involve $K$, the corresponding results remain unchanged with different $K$.

In the simulated \quotes{species} dataset (Figure~\ref{fig:manualK}(a)-(c)), there are several noticeable observations. First, the recall obtained by COCACOLA maintains around $\sim 0.999$, as stably as plateau, slightly larger than $\sim 0.996$ by CONCOCT and $0.9973$ by MaxBin, much higher than $0.881$ by GroopM and $0.9174$ by MetaBAT. The nearly perfect recall for COCACOLA, CONCOCT and MaxBin suggests that almost all contigs in each species are grouped within the same cluster no matter how $K$ is chosen. Second, the precision of both COCACOLA and CONCOCT improves as $K$ increases with some deviations. When $K$ is small, some clusters may contain contigs from different species. As $K$ increases, the contigs in each bin tend to be more homogeneous that they come from the same species, thus precision improves. Notice that MetaBAT performs very well in terms of precision and ARI, it requires a large $K$ for COCACOLA to catch up with.

In the simulated \quotes{strain} dataset (Figure~\ref{fig:manualK}(d)-(f)), when the number of clusters $K$ varies from $20$ to $80$, COCACOLA reaches a plateau at $K=26$ with respect to precision, recall and ARI. Specifically, precision stabilizes at $\sim 0.976$. Recall stabilizes at $\sim 0.976$ meanwhile ARI stabilizes at $\sim 0.9516$. COCACOLA outperforms CONCOCT, MaxBin and MetaBAT almost in all cases except when $K=78$.

We conclude that COCACOLA tends to perform better with respect to larger $K$ than the exact number the real species, it is understandable because larger $K$ explores the data better.

\subsection{The Effect of the Number of Samples on the Performance}
To evaluate the effect of the number of samples on the performance, we use a fraction of samples as input only. The simulated \quotes{species} dataset comprises $96$ samples overall. Thus we choose sub-samples of size ranging from $10$ to $90$, with $10$ as increment. To avoid duplicate contribution from a particular sample, we choose sub-samples without overlapping. Therefore, the numbers of sub-samples are $9,4,3,2,1,1,1,1,1$, respectively.

\begin{figure}
\centering
\begin{tabular}{ccc}
\includegraphics[width=0.32\textwidth]{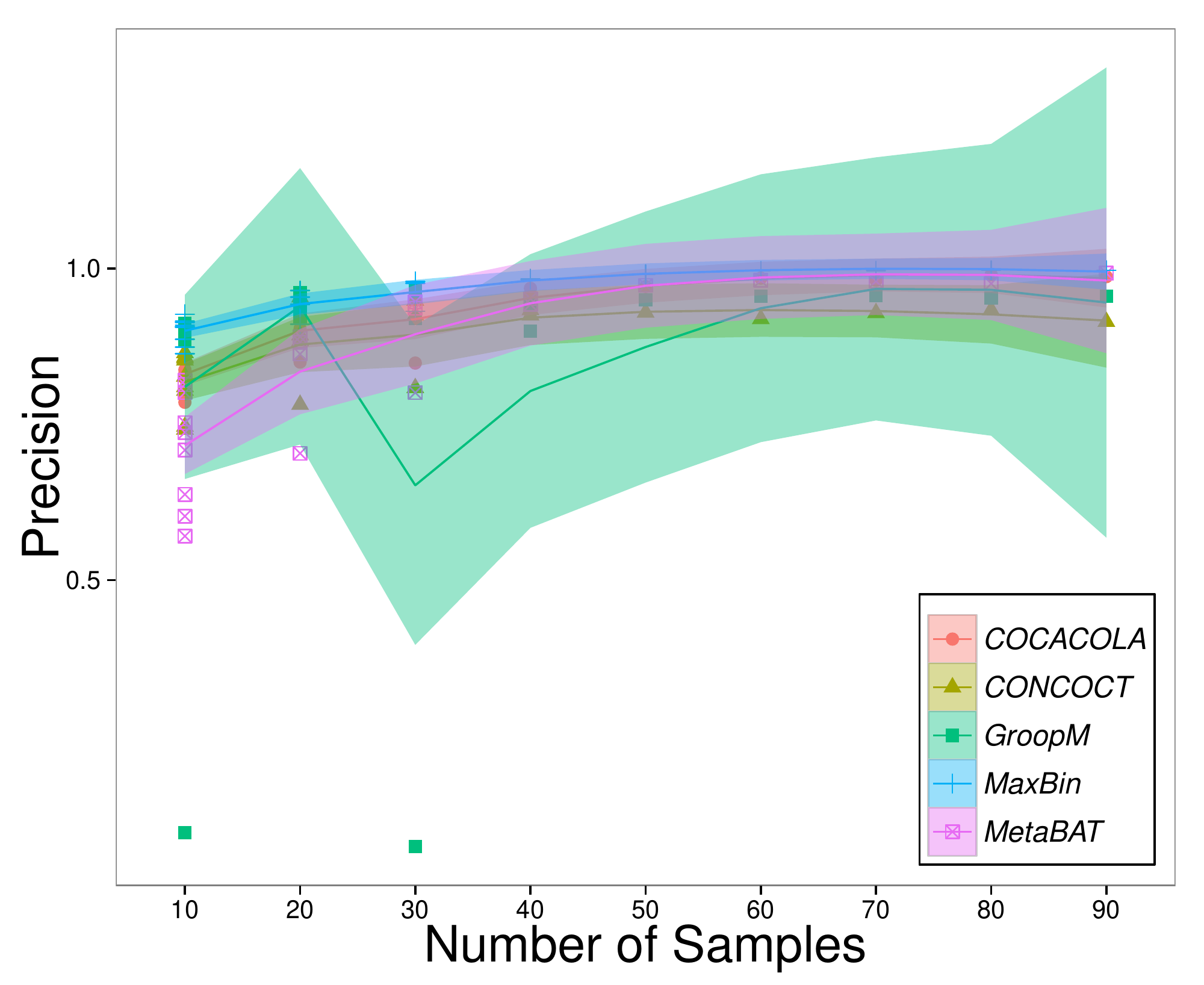} &
\includegraphics[width=0.32\textwidth]{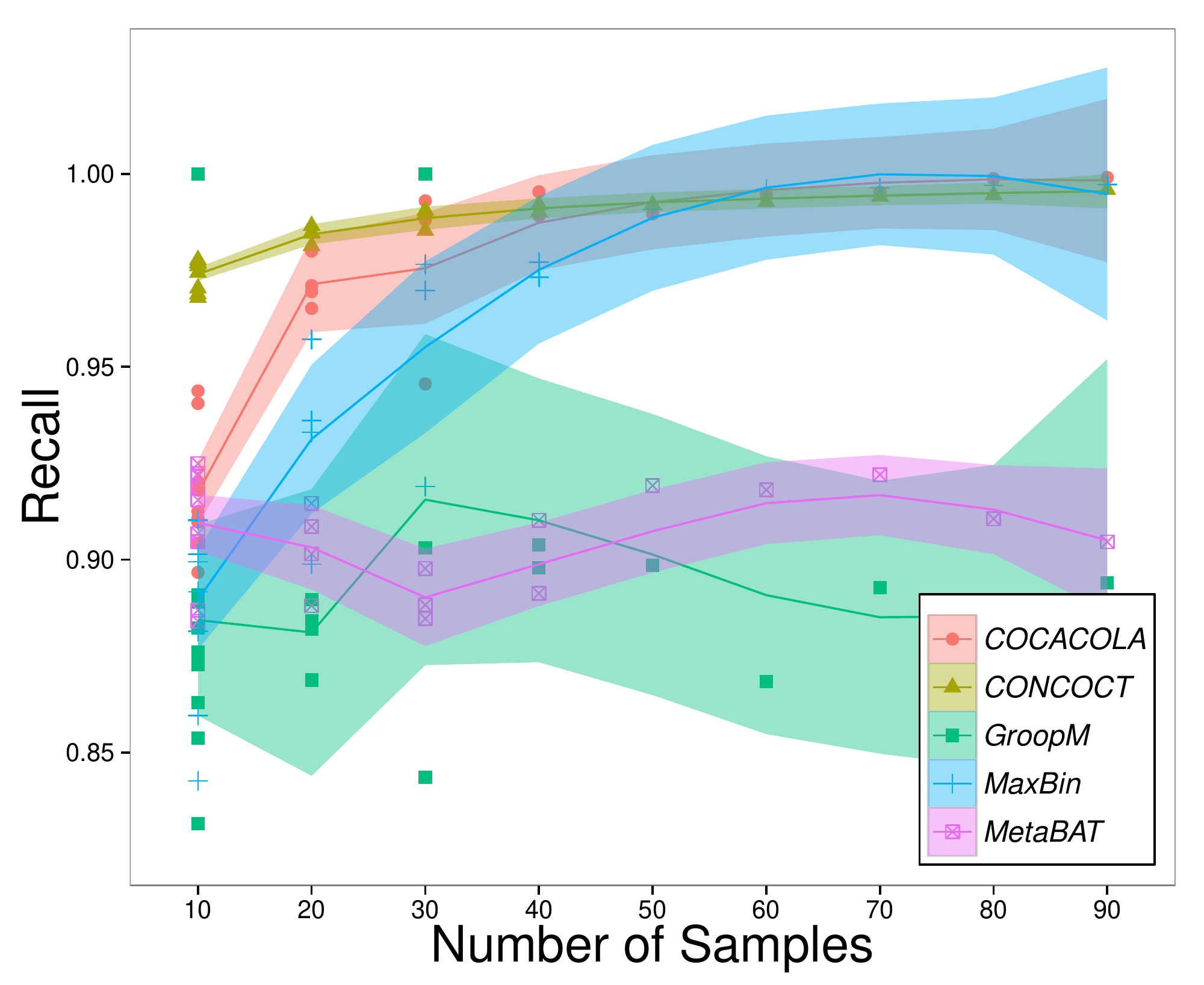} &
\includegraphics[width=0.32\textwidth]{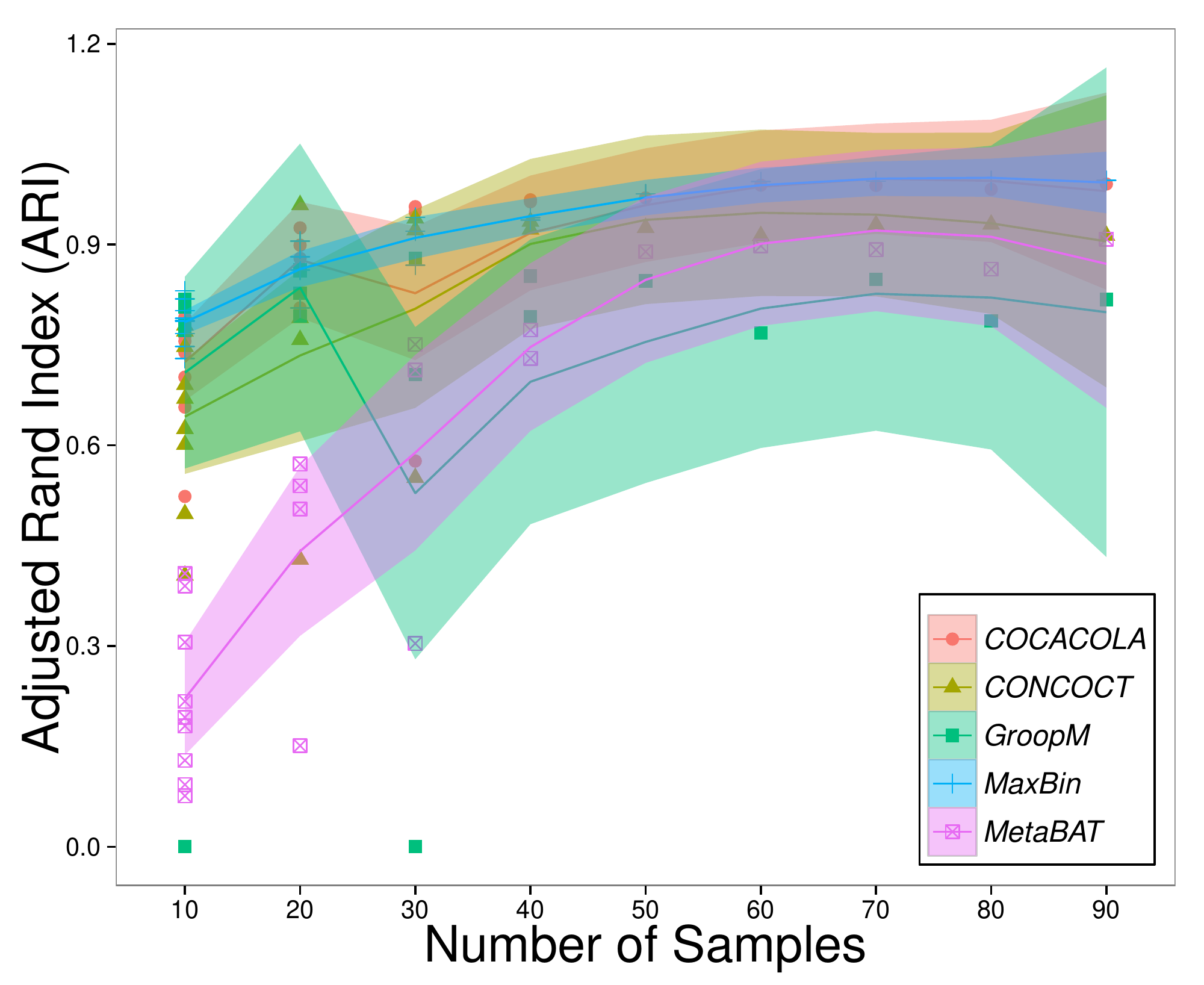} \\
(a) & (b) & (c) \\
\end{tabular}
\caption{Evaluation of COCACOLA, CONCOCT, GroopM, MaxBin and MetaBAT on sensitivity to varying sample size on simulated \quotes{species} dataset. The number of sub-samples are 9,4,3,2,1,1,1,1,1, corresponding to sample size 10,20,30,40,50,60,70,80,90, respectively. The result is smoothed by Local Polynomial Regression with confidence interval.}
\label{fig:speciesSubsamp}
\end{figure}

We set the number of clusters $K=200$ for all sub-sample studies. Fig  \ref{fig:speciesSubsamp} shows the precision, recall and ARI of five methods for the different sub-samples. Notice that the precision, recall and ARI start to decrease when sample size drops below $K=40$. The major contribution to the decrease is due to the loss of recall, interpreted as over-estimation of OTU numbers. The reasons lie in the fact that hard clustering-like approaches such as COCACOLA are less robust compared to mixture model-like approaches such as CONCOCT. Despite that, COCACOLA demonstrates more robustness in terms of precision and ARI than CONCOCT across all sub-samples.

GroopM demonstrates unstable performance when sample size is small. For instance, as depicted in Figure~\ref{fig:speciesSubsamp}(a) and Figure~\ref{fig:speciesSubsamp}(c), GroopM groups all contigs into one wholistic bin in one sub-sample of size $10$ and $30$, respectively. In comparison, the bin sizes obtained  are $84$ and $96$ by COCACOLA, $80$ and $92$ by CONCOCT, respectively.

MaxBin performs the best in the case of small sample size with regard to precision and ARI, however, the superiority roots in the fact that MaxBin excludes a proportion of ambiguous contigs for binning. In comparison, rest methods consider all contigs.

We conclude that COCACOLA and CONCOCT turn out to perform more stably in the case of small sample size compared to GroopM and MetaBAT. When sample size is small, COCACOLA and CONCOCT display different trade-offs between precision and recall. In particular, COCACOLA shows superiority in terms of precision whereas inferiority in terms of recall in comparison to CONCOCT. Overall, COCACOLA outperforms CONCOCT in terms of ARI, an indicator taking both precision and recall into account. When sample size grows large, COCACOLA outperforms CONCOCT in all respects.

\subsection{The Effect of Incorporating Additional Information on Binning}
We investigate the performance improvement of COCACOLA after incorporating two additional knowledge as proposed in the \quotes{Methods} section, in particular, co-alignment to reference genomes and linkage between contigs provided by paired-end reads. Moreover, we study the ensemble of both. The comparison is between the binning result by COCACOLA incorporating additional knowledge against the result without. The comparison is based upon sub-samples of the simulated \quotes{species} dataset. Since the contributions from additional knowledge nearly diminish when the sample size exceeds $K=30$, therefore we focus on the $16$ cases from $K=10$ to $K=30$.

\begin{figure}
\centering
\begin{tabular}{ccc}
\includegraphics[width=0.3\textwidth]{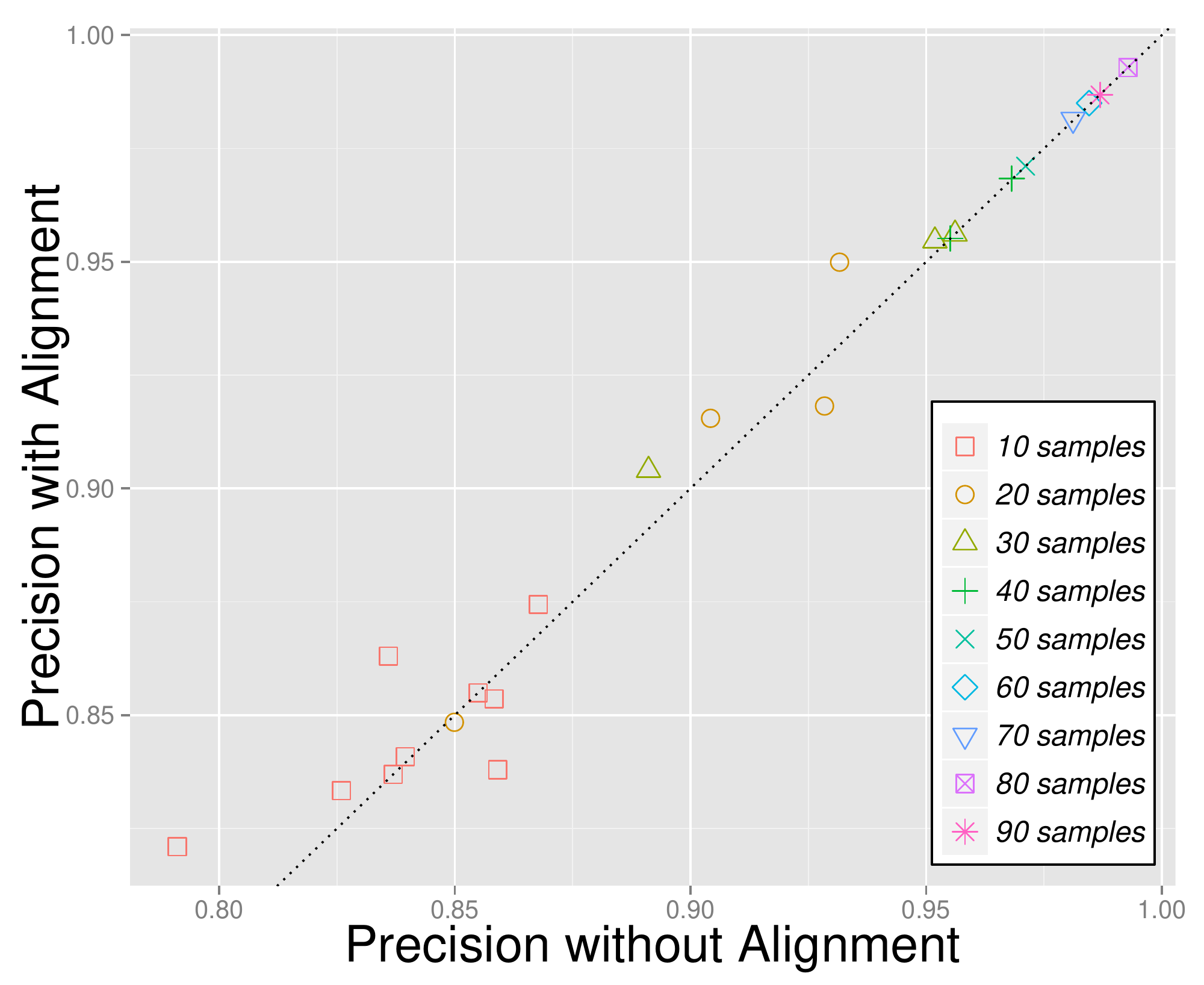} &
\includegraphics[width=0.3\textwidth]{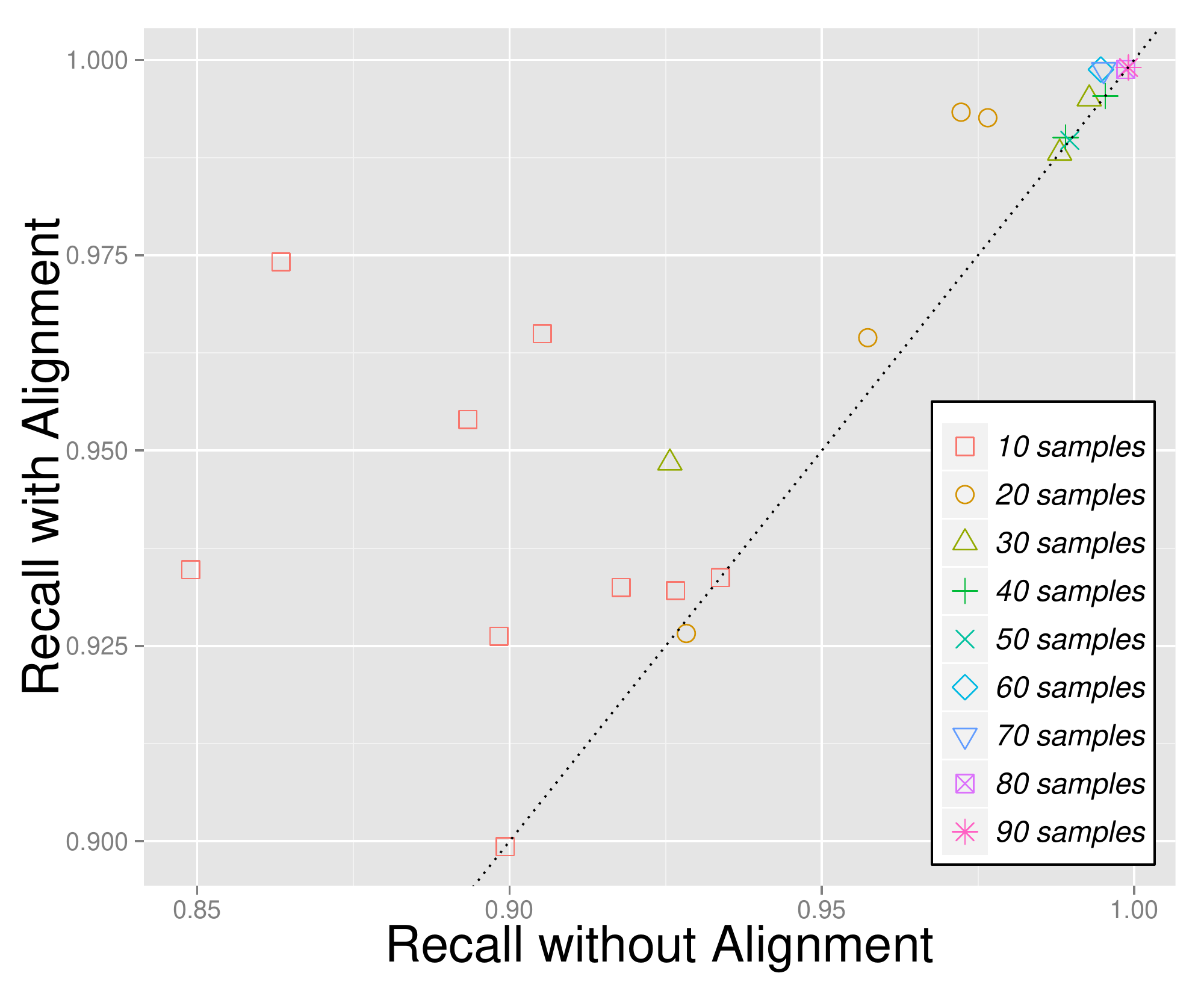} &
\includegraphics[width=0.3\textwidth]{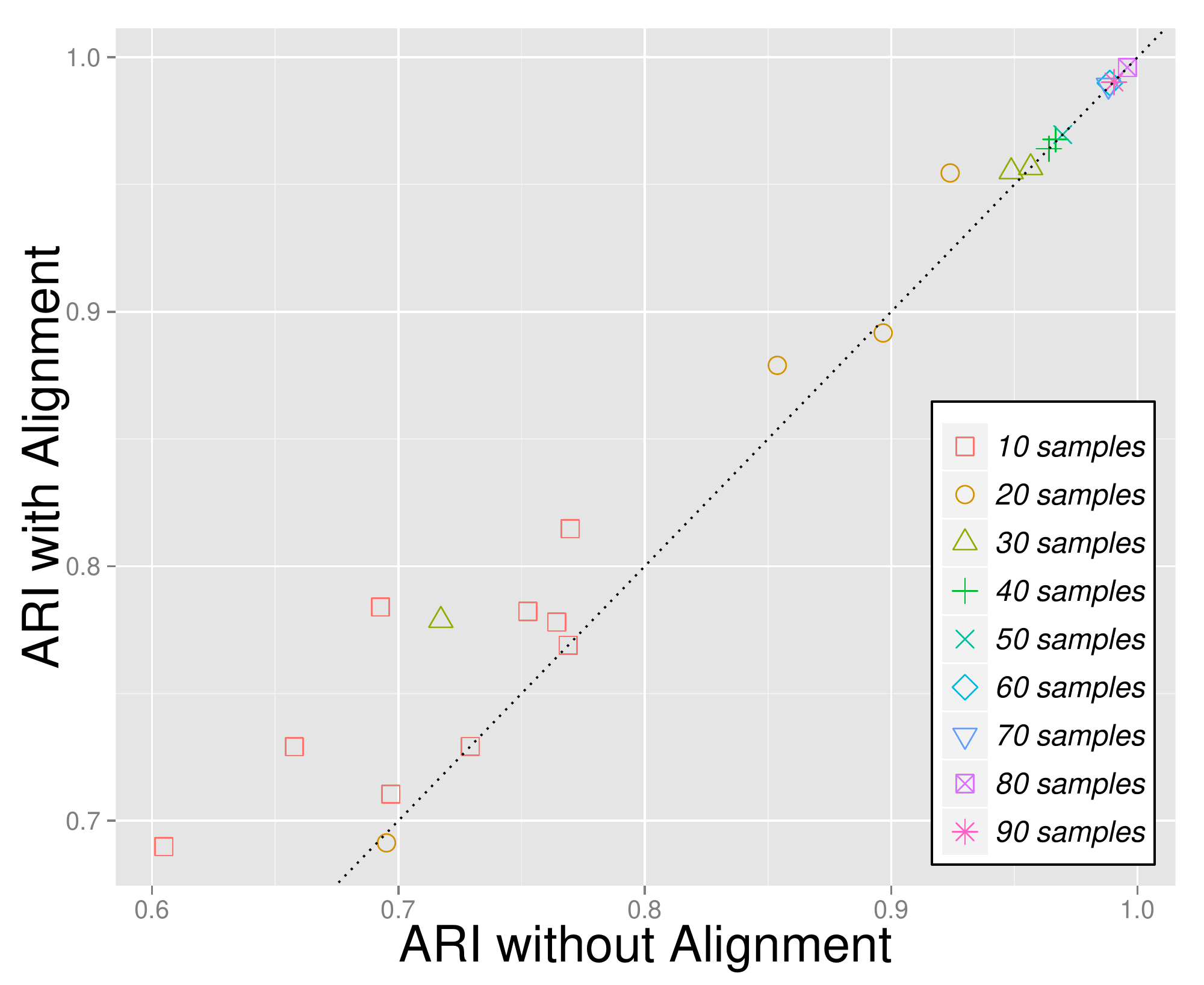} \\
(a)  & (b)  & (c)  \\
\includegraphics[width=0.3\textwidth]{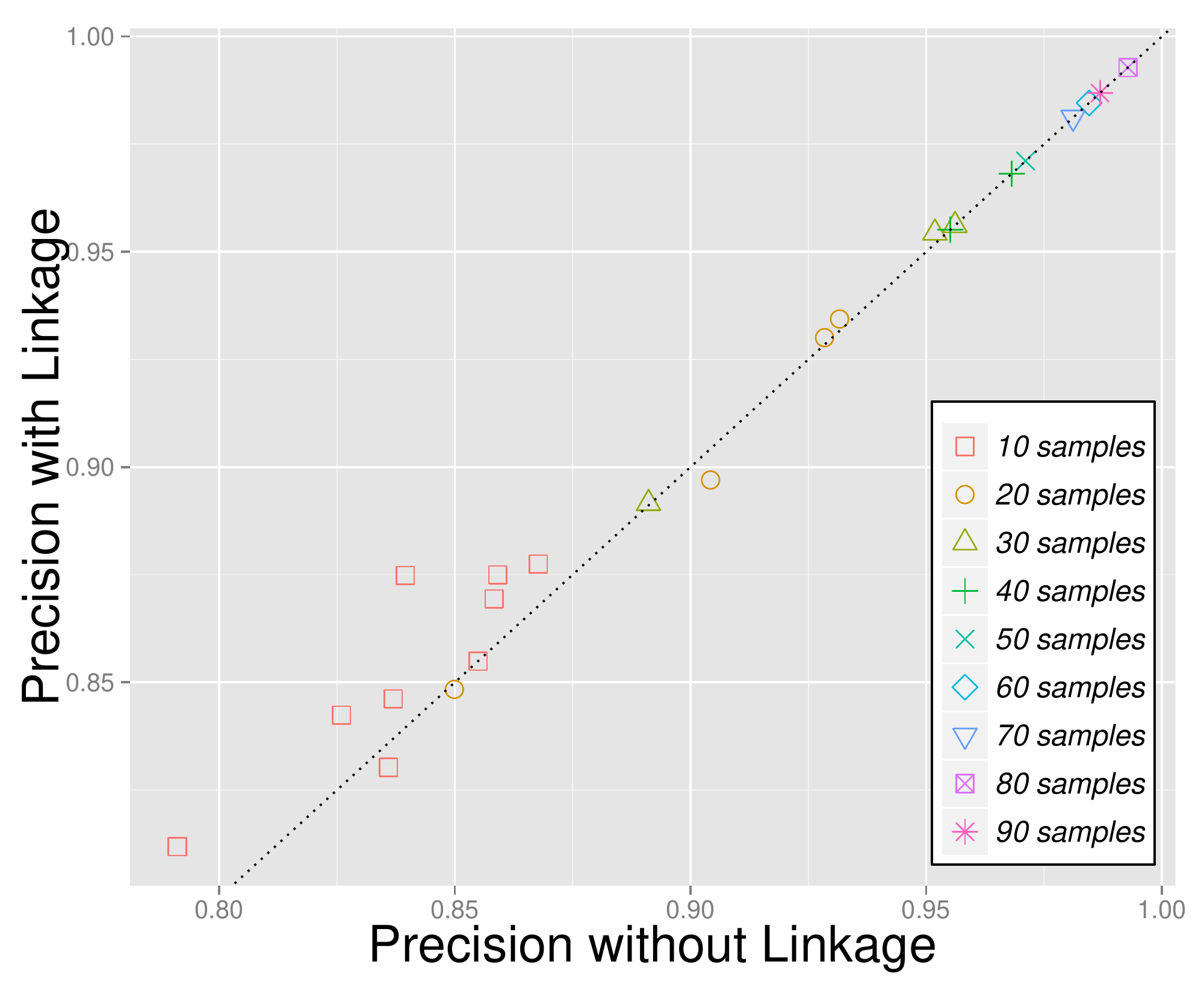} &
\includegraphics[width=0.3\textwidth]{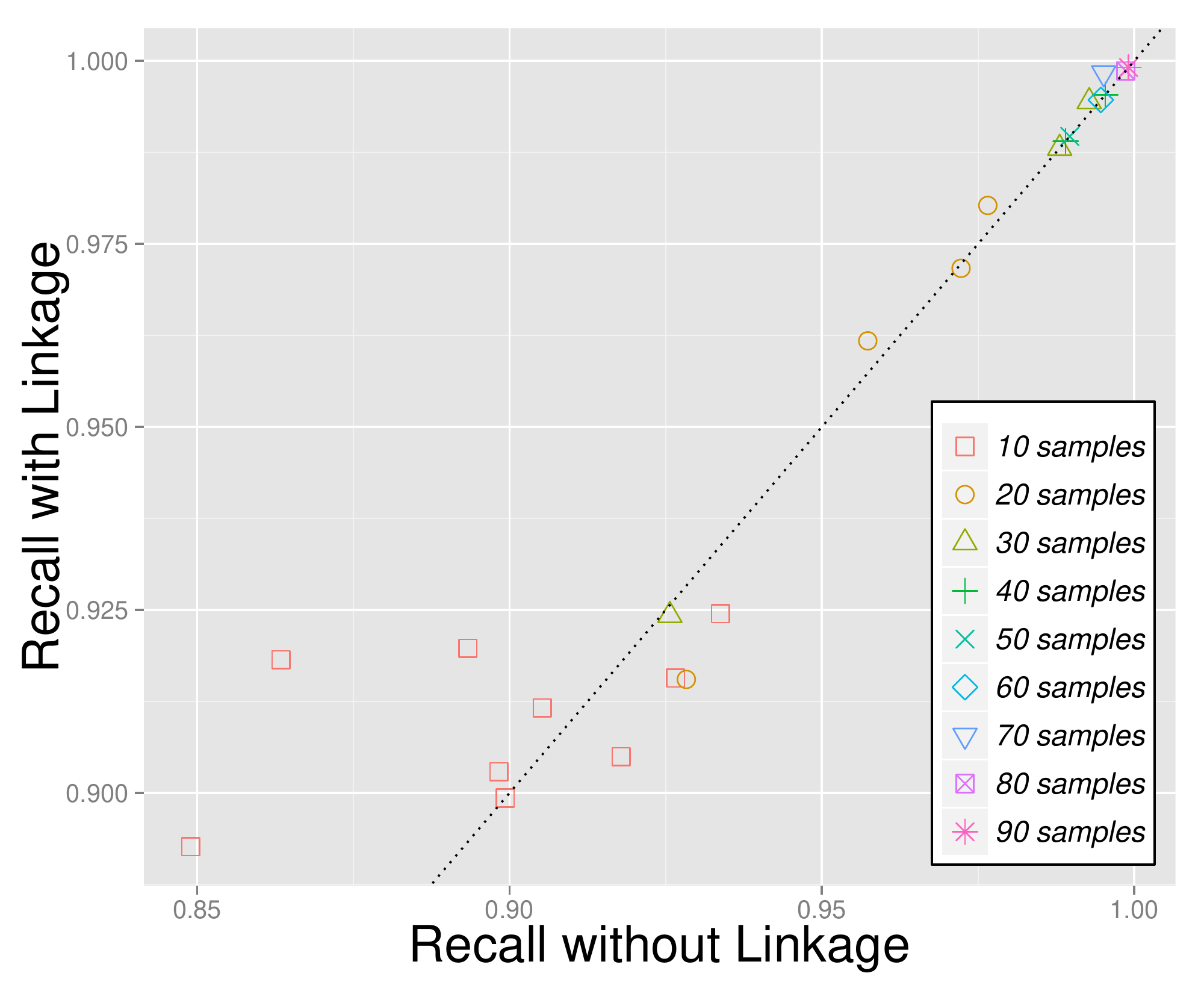} &
\includegraphics[width=0.3\textwidth]{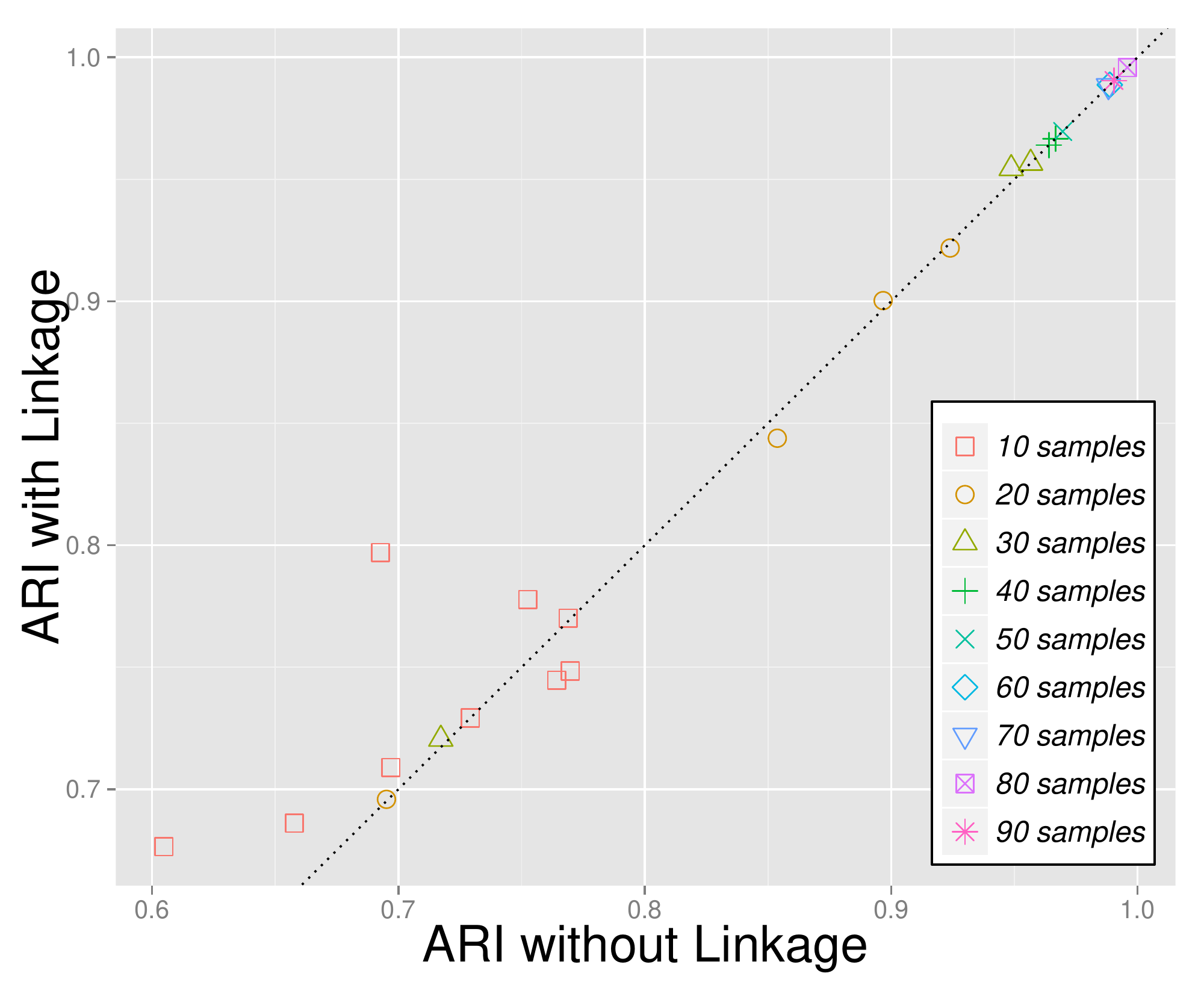} \\
(d)  & (e)  & (f)  \\
\includegraphics[width=0.3\textwidth]{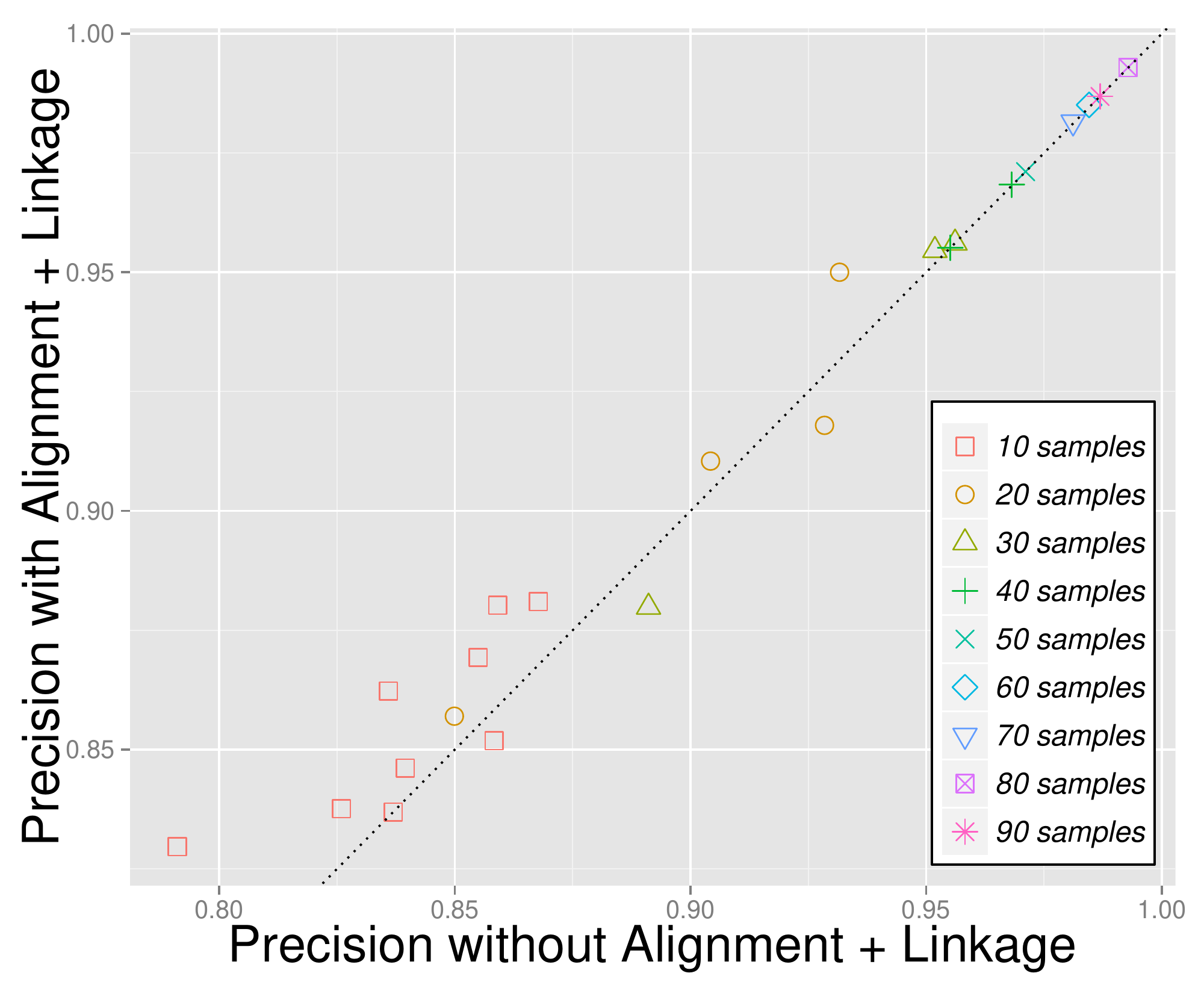} &
\includegraphics[width=0.3\textwidth]{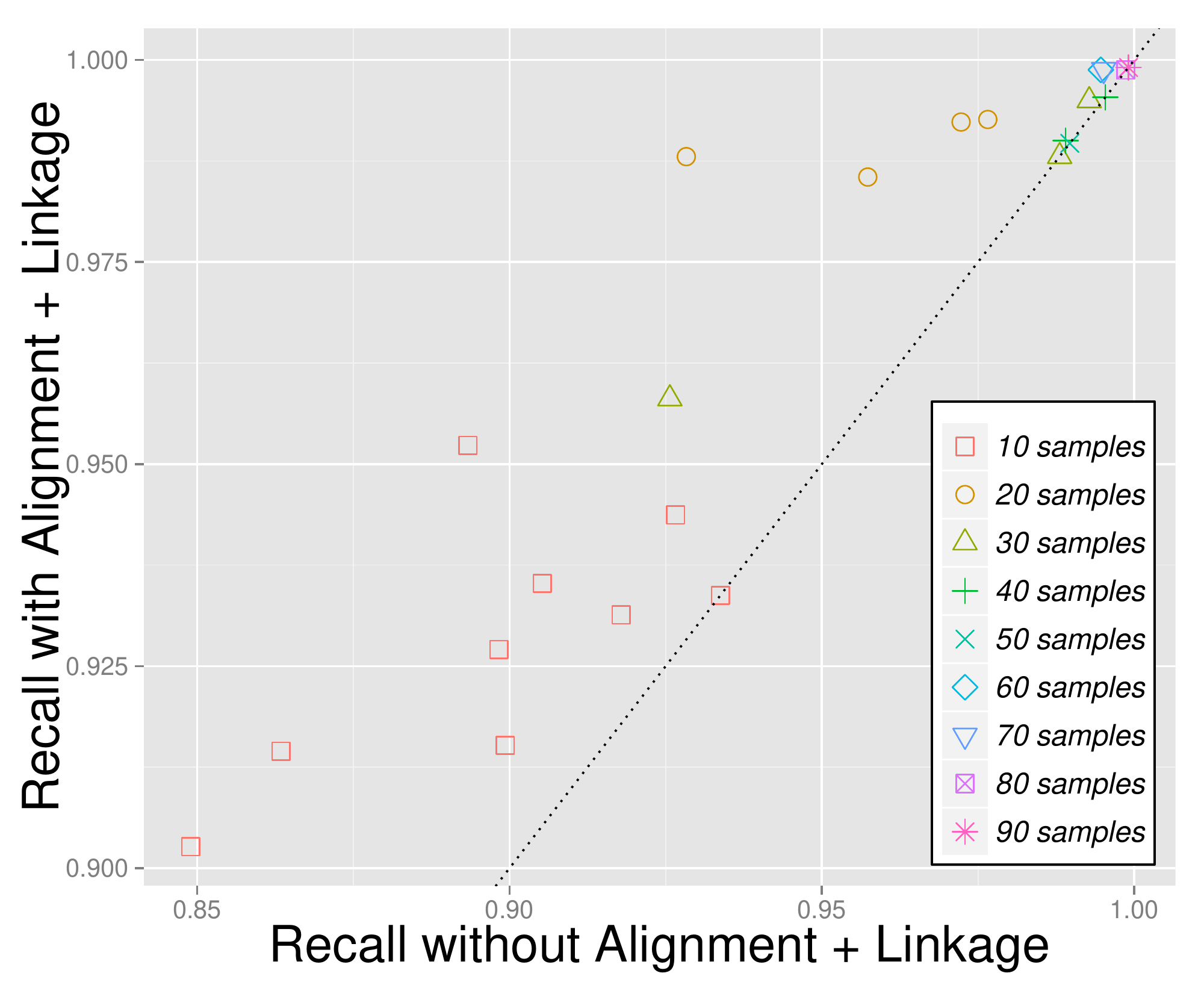} &
\includegraphics[width=0.3\textwidth]{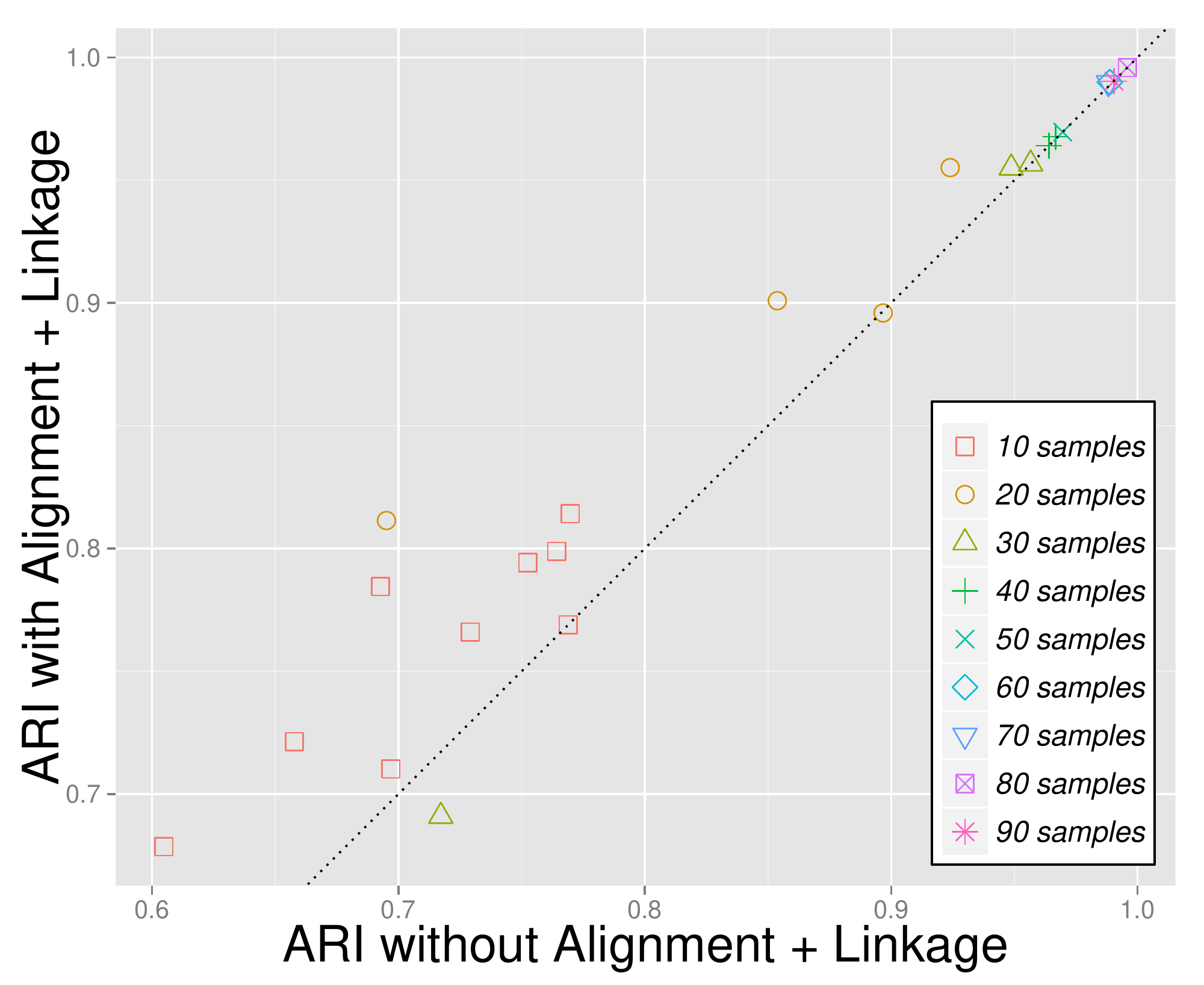} \\
(g)  & (h)  & (i)  \\
\end{tabular}
\caption{Evaluation of the impact of incorporating two additional knowledge (Section \ref{additionInfo}) on sub-samples of simulated \quotes{species} dataset. The first option is co-alignment information to reference genomes, depicted by (a)-(c). The second option is paired-end reads linkage, depicted by (d)-(f). The ensemble of both is depicted by (g)-(i).}
\label{fig:speciesSubsampAddInfo}
\end{figure}

In terms of co-alignment, we design the symmetric weight matrix $\mathbf{A}_{n n'}=1$ if contig $n$ and contig $n'$ are aligned to the same species using the TAXAassign script \cite{taxaassign}. As shown in Figure~\ref{fig:speciesSubsampAddInfo}(a)-(c), the precision is improved noticeably in $7$ cases and decreased in $3$ cases, the recall is improved noticeably in $11$ cases and decreased slightly in $1$ case, the ARI is improved noticeably in $10$ cases and decreased slightly in $2$ case.

In terms of linkage, we design the symmetric weighted matrix $\mathbf{A}_{n n'}$ as the number of samples supporting linkage connecting contig $n$ and contig $n'$. As depicted in Figure~\ref{fig:speciesSubsampAddInfo}(d)-(f), the precision is improved noticeably in $7$ cases and decreased in $2$ cases, the recall is improved noticeably in $7$ cases and decreased slightly in $4$ cases, the ARI is improved noticeably in $5$ cases and decreased in $3$ cases.

In terms of the ensemble of co-alignment and linkage, as depicted in Figure~\ref{fig:speciesSubsampAddInfo}(g)-(i), the precision is improved noticeably in $10$ cases and decreased in $3$ cases, the recall is improved noticeably in $13$ cases and no case suffers decreasing, the ARI is improved noticeably in $11$ cases and decreased in $1$ cases.

We have the following conclusions: (1) When there are sufficient number of samples, the contributions from additional knowledge diminish. (2) Additional knowledge such as co-alignment and linkage information facilitate better overall performance in the majority of cases. (3) Ensemble of both information performs more stable than individual information.
\subsection{Performance on Real Datasets}
Applying COCACOLA to the \quotes{Sharon} dataset (Figure~\ref{fig:autoK}(c)), given initial choice of $K=30$, the precision, recall and ARI reach $0.9889$, $0.9759$ and $0.9670$, respectively. In comparison, CONCOCT, GroopM, MaxBin and MetaBAT achieve $0.9801$, $0.9820$, $0.7077$ and $0.9705$ in terms of precision, $0.9606$, $0.9147$, $0.9767$ and $0.8344$ in terms of recall, $0.9600$, $0.9126$, $0.5639$ and $0.8634$ in terms of ARI, respectively. COCACOLA identifies $6$ OTUs corresponding to six reported genomes. In comparison, CONCOCT, GroopM, MaxBin and MetaBAT identify $14$, $24$, $5$ and $11$ OTUs, respectively.

Next we investigate the performance improvement of COCACOLA after incorporating additional knowledge. We use linkage information only because it's circular to use TAXAassign script \cite{taxaassign} on both alignment and labeling. COCACOLA still identifies 6 OTUs, with the precision, recall and ARI reaching $0.9923$, $0.9797$ and $0.9743$, slightly outperforms the case without additional knowledge.

Applying COCACOLA to the \quotes{MetaHIT} dataset (Figure~\ref{fig:autoK}(d)), given initial choice of $K=100$, the precision, recall and ARI reach $0.9082$, $0.8272$ and $0.7717$, respectively. In comparison, CONCOCT, GroopM, MaxBin and MetaBAT achieve $0.8933$, $0.5247$, $0.6655$ and $0.5738$ in terms of precision, $0.7901$, $0.6843$, $0.8228$ and $0.7397$ in terms of recall, $0.7518$, $0.3757$, $0.5866$ and $0.1088$ in terms of ARI, respectively.

Next we investigate the performance improvement of COCACOLA after incorporating linkage information. The performance is further slightly improved from $0.9082$ to $0.9084$ in terms of precision, from $0.8272$ to $0.8350$ in terms of recall, and from $0.7717$ to $0.7844$ in terms of ARI, respectively.
\subsection{Running Time of COCACOLA, CONCOCT, GroopM, MaxBin and MetaBAT}
COCACOLA shares the same data parsing pipeline as CONCOCT and differs only in the binning step, whereas GroopM uses its own workflow. It is reasonable to compare running time of binning directly between COCACOLA and CONCOCT. In order to bring GroopM into context, we take into account the stages related to binning and therefore exclude the data parse stage. As for MaxBin and MetaBAT we simply pre-calculate the abundance and depth information. MaxBin involves multi-threaded parameter, which is set as the number of cores. All of five methods run on the 12-cores and 60GB-RAM computing platform provided by the USC High Performance Computing Cluster. The comparison is conducted on both the simulated datasets and real datasets (Table~\ref{tab:time}).
We conclude that COCACOLA runs faster than CONCOCT, GroopM, MaxBin and MetaBAT.

\begin{table}
\centering
\small
\begin{tabular}{|c|c|c|c|c|c|c|c|c|c|c|}
\hline
\multirow{2}{*}{Dataset} &\multicolumn{2}{c|}{COCACOLA} &\multicolumn{2}{c|}{CONCOCT} &\multicolumn{2}{c|}{GroopM} &\multicolumn{2}{c|}{MaxBin} &\multicolumn{2}{c|}{MetaBAT}\\
\cline{2-11}
 & Time & Speedup & Time & Speedup & Time & Speedup & Time & Speedup & Time & Speedup \\
\hline
\quotes{species} & 1m41.50s & 1x & 17m14.71s & 10.2x & 1h57m28s & 69.4x & 49m48.52s & 29.4x & 4m16.14s & 2.5x\\
\quotes{strain} & 10.94s & 1x & 1m10.99s & 6.5x & 17m00.46s & 93.3x & 9m54.80s & 54.4x & 2m31.52s & 13.9x\\
\quotes{Sharon} & 13.22s & 1x & 25.11s & 1.9x & 4m45.85s & 21.6x & 1m36.09s & 7.3x & 24.66s & 1.9x \\
\quotes{MetaHIT} & 2m39.12s & 1x & 20m20.90s & 7.7x & 12m47.68s & 4.8x & 2h20m52s & 53.1x & 7m25.07s & 2.8x\\
\hline
\end{tabular}
\caption{\label{tab:time} Running Time of COCACOLA, CONCOCT, GroopM, MaxBin and MetaBAT}
\end{table}

\subsection{The Effect of Initialization}
To evaluate the effect of initialization on the performance, we conduct the comparison $L_{1}$ distance and $L_{2}$ (Euclidean) distance on sub-samples of simulated \quotes{species} dataset.

\begin{figure}
\centering
\begin{tabular}{ccc}
\includegraphics[width=0.32\textwidth]{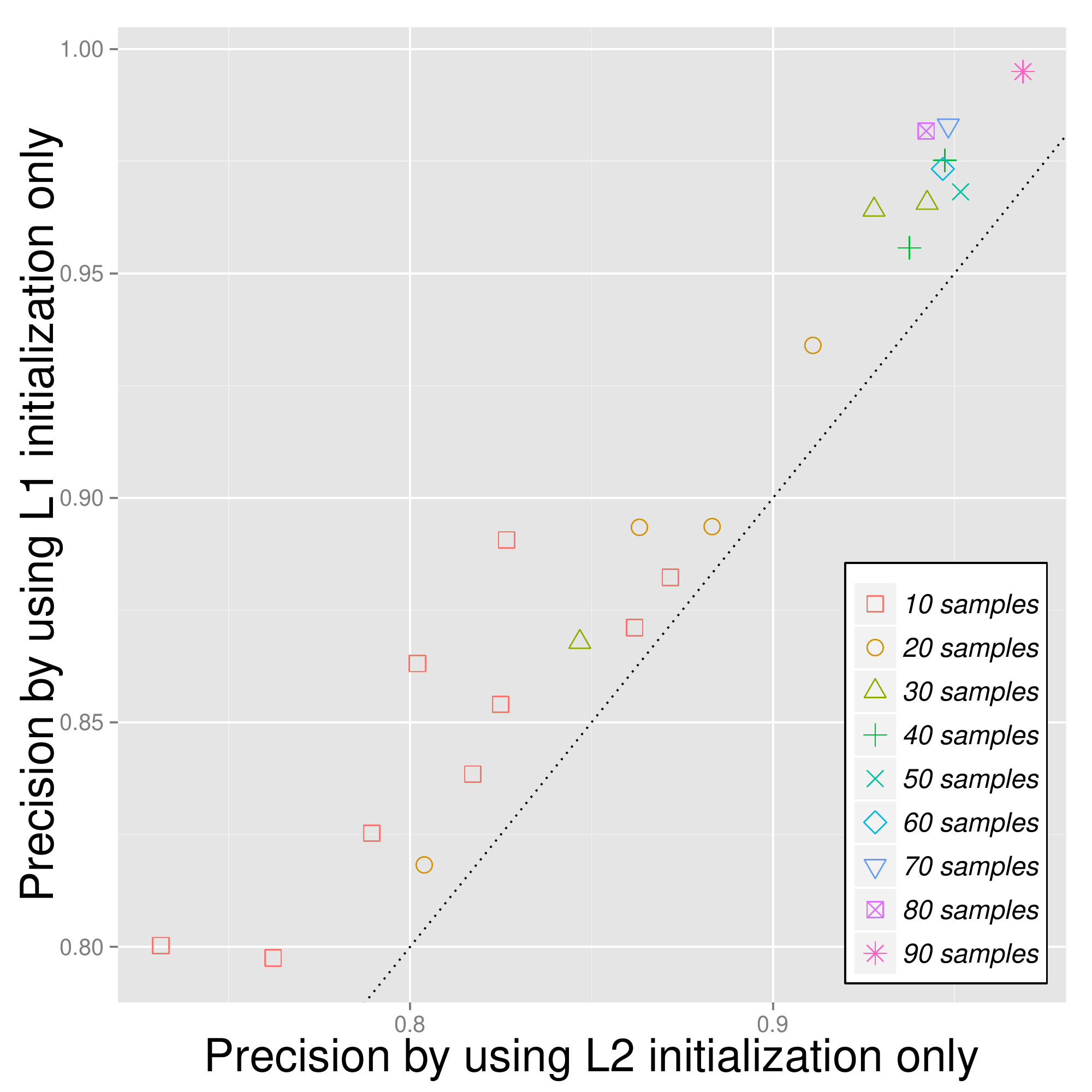} &
\includegraphics[width=0.32\textwidth]{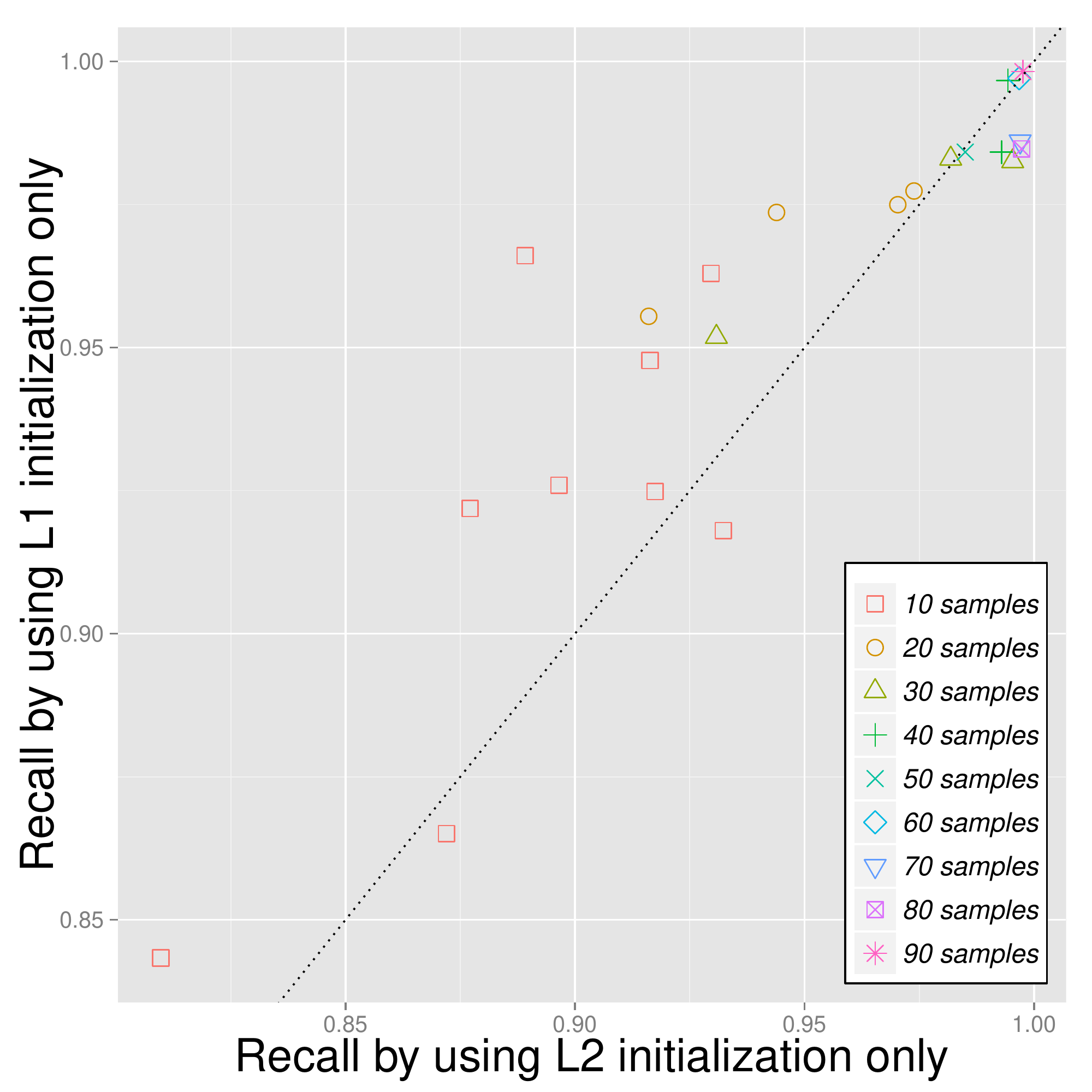} &
\includegraphics[width=0.32\textwidth]{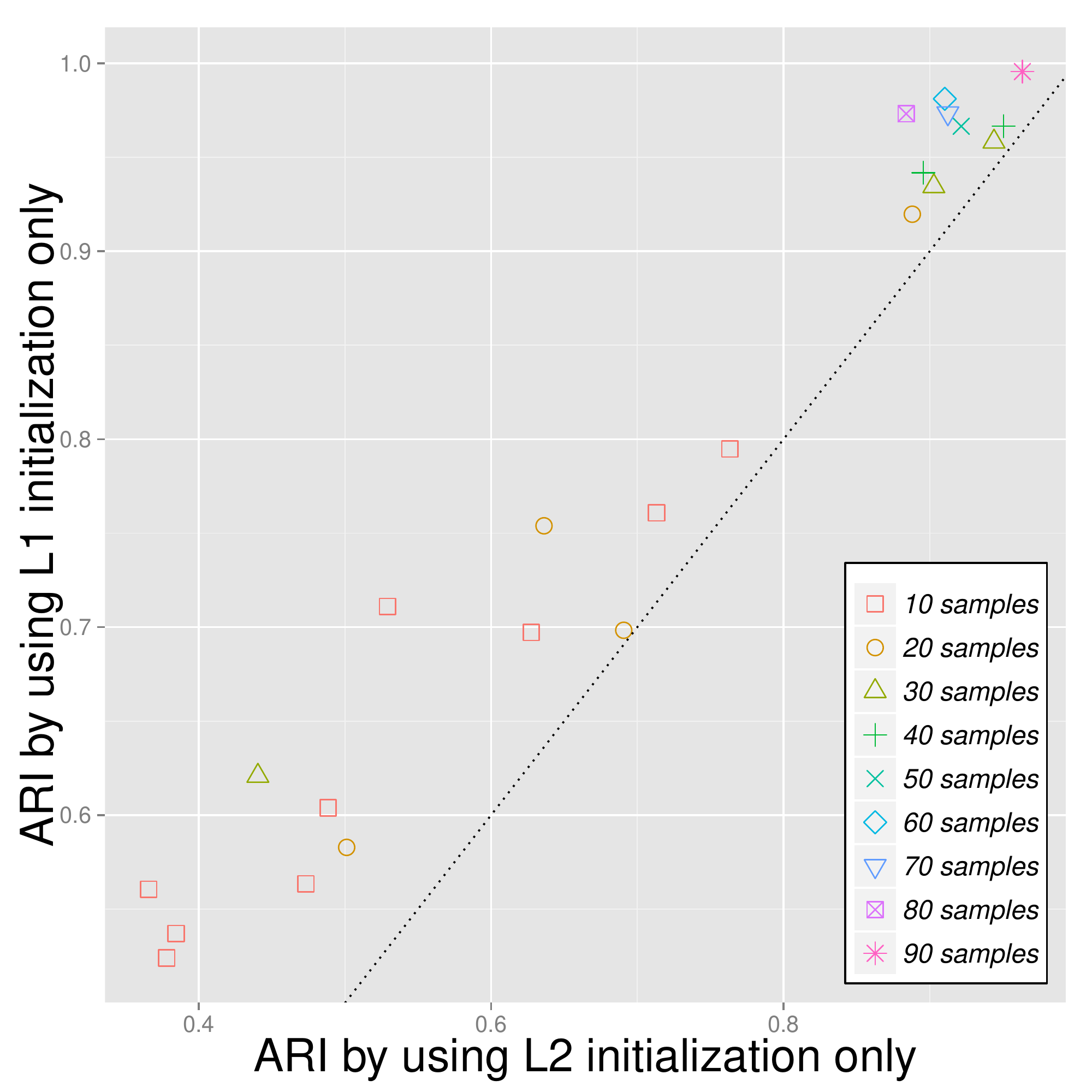} \\
(a)  & (b)  & (c)  \\[6pt]
\includegraphics[width=0.32\textwidth]{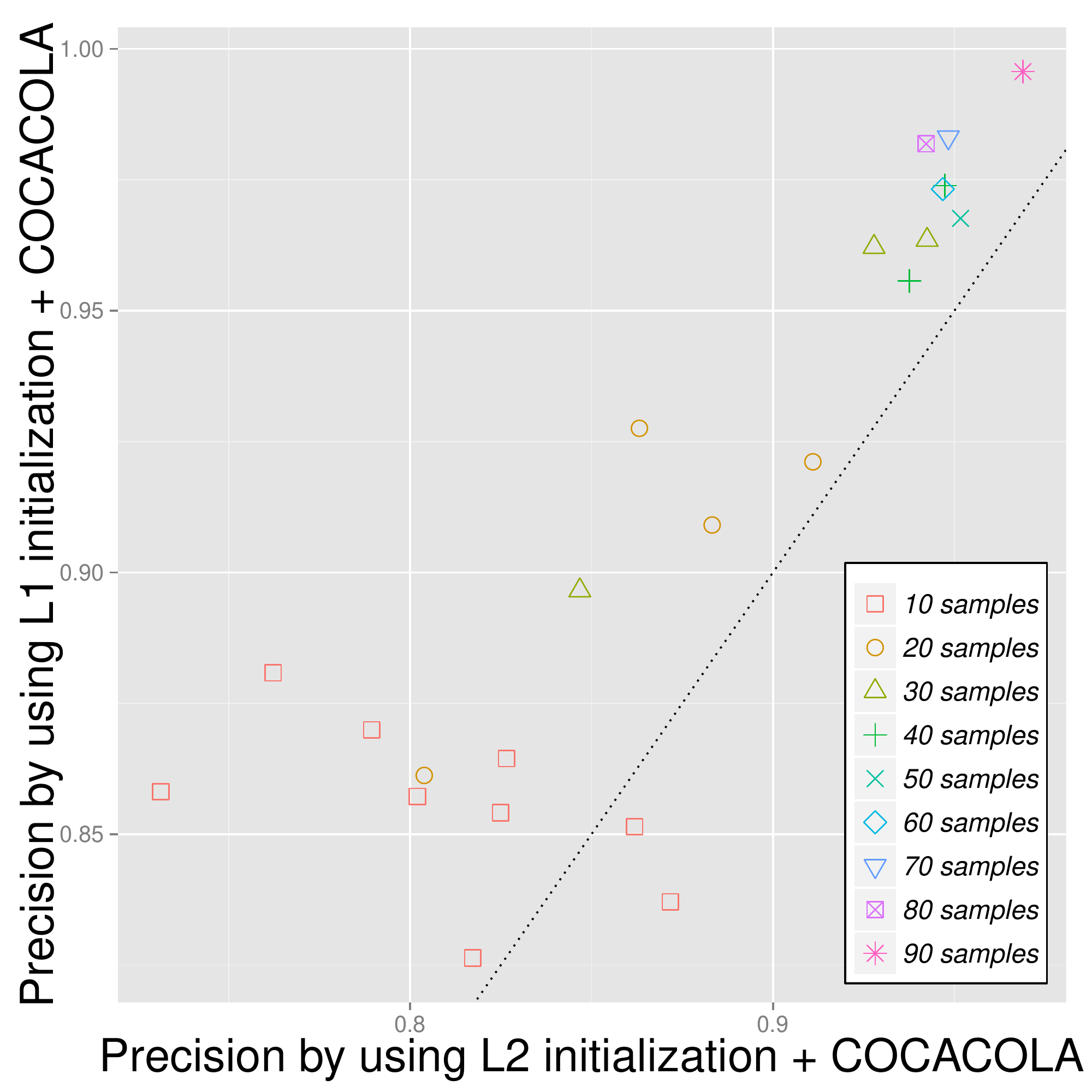} &
\includegraphics[width=0.32\textwidth]{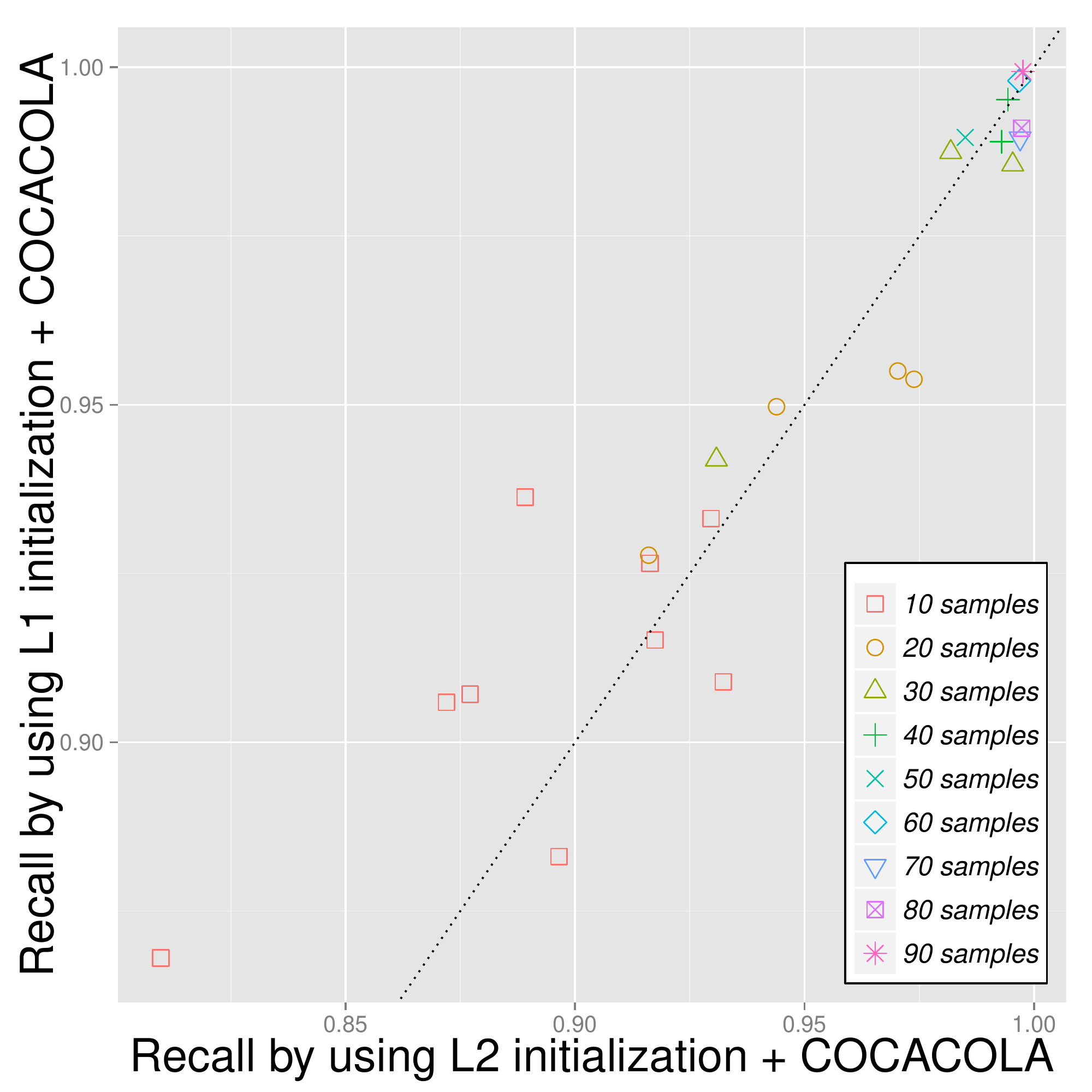} &
\includegraphics[width=0.32\textwidth]{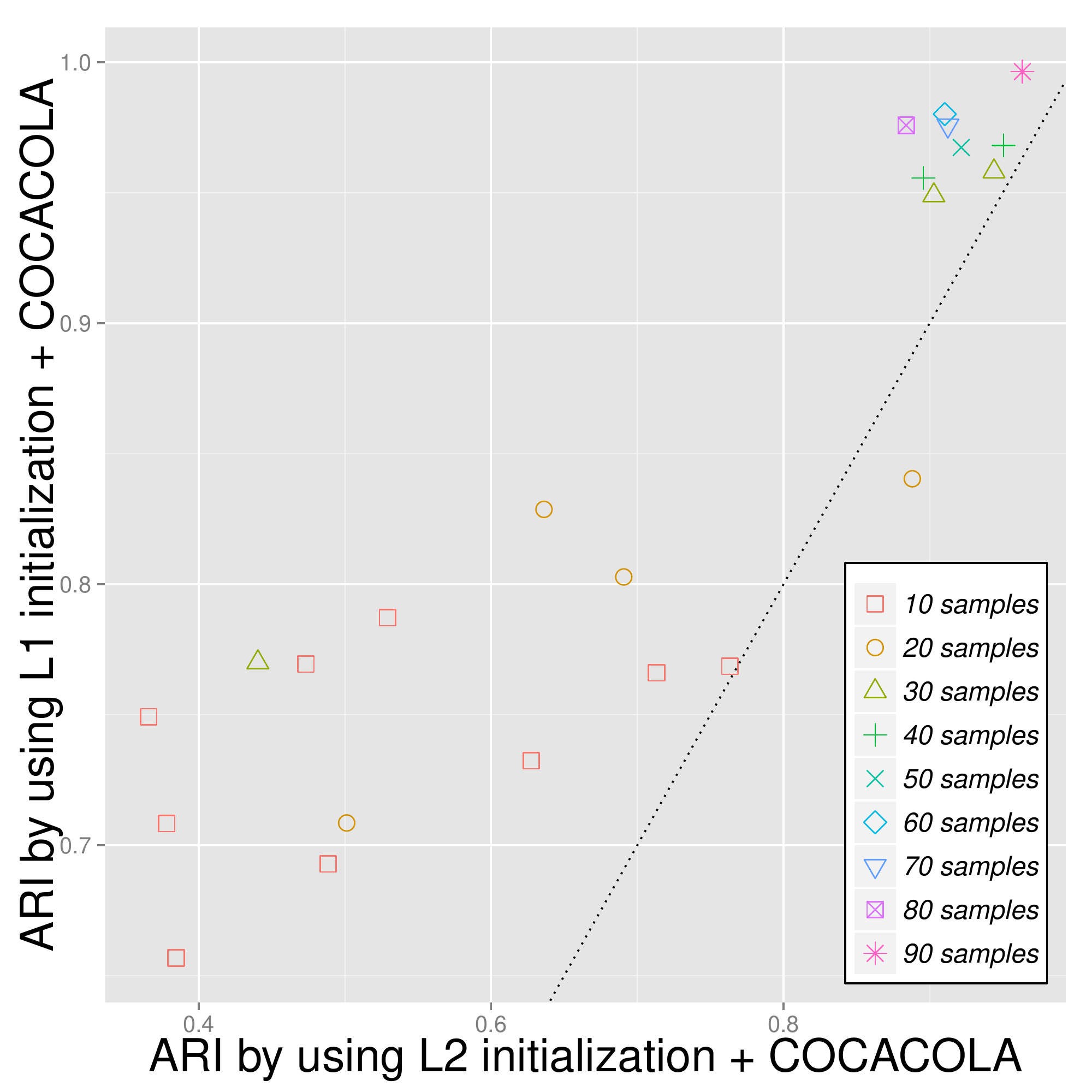} \\
(d)  & (e)  & (f)  \\[6pt]
\end{tabular}
\caption{Evaluation of the impact of initialization on sub-samples of simulated \quotes{species} dataset. The performance comparison only based on initialization with different distance measurements are depicted by (a)-(c). The performance comparison based on initialization with different distance measurements and COCACOLA are depicted by (d)-(f).}
\label{fig:speciesSubsampCompInit}
\end{figure}

We first evaluate the performance originating from initialization only (Figure~\ref{fig:speciesSubsampCompInit}(a)-(c)). In $6$ out of $23$ cases, initialization using $L_{2}$ distance shows better precision whereas suffers worse recall. In the rest cases, initialization using $L_{2}$ distance is dominated by $L_{1}$ distance.

We then evaluate the performance of COCACOLA with initialization using $L_{1}$ and $L_{2}$ distance. In $8$ out of $23$ cases, initialization using $L_{2}$ distance shows better precision whereas suffers worse recall. And in $2$ out of $23$ cases, initialization using $L_{2}$ distance shows better recall whereas suffers worse precision. In the rest cases, initialization using $L_{2}$ distance is dominated by $L_{1}$ distance. As for ARI, initialization using $L_{2}$ distance outperforms $L_{1}$ distance only in $1$ out of $23$ cases.

We conclude that $L_{1}$ distance performs better than $L_{2}$ (Euclidean) distance in initialization.

\section{Discussion}
\label{discussion}
In this paper, we develop a general framework to bin metagenomic contigs utilizing sequence
composition and coverage across multiple samples. Our approach, COCACOLA, outperforms state-of-art binning approaches CONCOCT \cite{alneberg2014binning}, GroopM \cite{imelfort2014groopm}, MaxBin \cite{wu2015maxbin} and MetaBAT \cite{kang2015metabat} on both simulated and real datasets.

The superior performance of COCACOLA relies on several aspects. First, initialization plays an important role in binning accuracy. Second, COCACOLA employs $L_{1}$ distance instead of Euclidean distance for better taxonomic identification. Third, COCACOLA takes advantage of both hard clustering and soft clustering. Specifically, soft clustering (such as the Gaussian mixture model used by CONCOCT) allows a contig to be assigned probabilistically to multiple OTUs, hence gains more robust results in general in comparison to hard clustering (such as the Hough partitioning used by GroopM). However, in complex environmental samples with strain-level variations, the corresponding OTUs are closely intertwined. Whereas soft clustering in turn further mixes the OTUs up and thus deteriorates clustering performance. COCACOLA obtains better trade-off between hard clustering and soft clustering by exploiting sparsity.

However, we notice that binning metagenomic contigs remains challenging when the number of samples is small, regardless of using COCACOLA, CONCOCT, GroopM, MaxBin or MetaBAT. With small number of metagenomic samples, the relationship between the contigs cannot be accurately inferred based on the relationship between the abundance profiles. Therefore, future research needs to study how to re-weight the contributions of abundance profiles and composition profiles in unsupervised \cite{cai2010unsupervised} or semi-supervised \cite{zhao2007semi} scenario. Moreover, recent studies suggest that Euclidean or $L_{1}$ distance between $l$-mer frequencies do not perform as well as alternative dissimilarity measurements such as $d_{2}^{*}$ and $d_{2}^{(shepp)}$ \cite{wan2010alignment} in comparing genome sequence. However, the use of such measurements is computationally challenging, which needs further exploration.

The COCACOLA framework seamlessly embraces customized knowledge to facilitate binning accuracy. In our study, we have investigated two types of knowledge, in particular, the co-alignment to reference genomes and linkage of contigs provided by paired-end reads. Even though the contributions from additional knowledge diminish when there are sufficient number of samples, they play an important role in binning results when the number of samples is small. In future studies, we intend to explore better customized prior knowledge. one option is exploiting phylogenetic information in taxonomic annotation \cite{purdom2011analysis}. Another option relies on identifying functional annotation of contigs, including open reading frames (ORF) that are likely to encode proteins \cite{ye2009orfome}, or co-abundance gene groups \cite{nielsen2014identification}, etc. We have also investigated the ensemble of both co-alignment and linkage knowledge, and it shows more stable performance than individual information. In future studies, we aim to find optimal weights \cite{tsuda2005fast} instead of equal weights.
\subsubsection*{Acknowledgments}

The authors would like to thank anonymous referees for helpful comments on this work. The
research is partially supported by NSF DMS-1043075 and OCE 1136818.

\bibliographystyle{splncs03}
\bibliography{proposal}

\end{document}